\documentclass[usenatbib]{mnras}
\usepackage{newtxtext,newtxmath}
\usepackage[T1]{fontenc}
\DeclareRobustCommand{\VAN}[3]{#2}
\let\VANthebibliography\thebibliography
\def\thebibliography{\DeclareRobustCommand{\VAN}[3]{##3}\VANthebibliography}
\usepackage{xcolor}

\usepackage{graphicx}	% Including figure files
\usepackage{tikz,tikz-3dplot}
\usetikzlibrary{3d,quotes,angles}
\usepackage[thinc]{esdiff}
\usepackage{braket}

\usepackage{amsmath}	% Advanced maths commands
\usepackage{mathtools}
\usepackage{hyperref}
\hypersetup{colorlinks=true,
            citecolor=blue,
            linkcolor=blue}
\usepackage[export,clipbox=false]{adjustbox}

\usepackage{chngcntr}
\usepackage{newtxtext,newtxmath}

\usepackage{varwidth}
\usepackage{soul}

\newcommand{\msun}{\mathrm{M}_{\odot}}	% Solar mass
\newcommand{\rh}{r_{\rm H}}

\newcommand{\mBH}{m_{\rm BH}}

\newcommand{\mSMBH}{M_\bullet}

\defcitealias{Whitehead2025}{W25b}

\defcitealias{Rowan2022}{Paper~I}
\defcitealias{Rowan2023}{Paper~II}
\begin{document}

%%%%%%%%%%%%%%%%%%% TITLE PAGE %%%%%%%%%%%%%%%%%%%

% Title of the paper, and the short title which is used in the headers.
% Keep the title short and informative.
\title[The alignment of disc transitors]{Hydrodynamic simulations of black hole evolution in AGN discs I: orbital alignment of highly inclined satellites}
% The list of authors, and the short list which is used in the headers.
% If you need two or more lines of authors, add an extra line using \newauthor
\author[C. Rowan et al.]{
Connar Rowan$^{1}$\thanks{E-mail: connar.rowan@nbi.ku.dk}, Henry Whitehead$^{2}$, Gaia Fabj$^{1}$, Philip Kirkeberg$^{1}$, Martin E. Pessah$^{1}$ and Bence Kocsis$^{3,4}$
\\
$^{1}$Niels Bohr International Academy, Niels Bohr Institute, Blegdamsvej 17, DK-2100 Copenhagen Ø, Denmark 
\\
$^{2}$Department of Physics, Astrophysics, University of Oxford, Denys Wilkinson Building, Keble Road, Oxford OX1 3RH, UK
\\
$^{3}$Rudolf Peierls Centre for Theoretical Physics, Clarendon Laboratory, University of Oxford, Parks Road, Oxford, OX1 3PU, UK
\\
$^{4}$St Hugh's College, St Margaret's Rd, Oxford, OX2 6LE, UK
}
% These dates will be filled out by the publisher
%\date{Accepted XXX. Received YYY; in original form ZZZ}

% Enter the current year, for the copyright statements etc.
%\pubyear{2015}

%\label{firstpage}
\date{\today}


\maketitle

\begin{abstract}
    The frequency of compact object interactions in AGN discs is naturally tied to the number of objects embedded within it. We investigate the evolution of black holes in the nuclear stellar cluster on inclined orbits to the AGN disc by performing adiabatic hydrodynamical simulations of isolated black hole disc crossings over a range of disc densities and inclinations $i\in[2^\circ,15^\circ]$. We find radiation dominates the pressure in the wake that forms around the BH across the full inclination and disc density range. We identify no well defined steady state wake morphology due to the thin geometry of the disc and the vertical exponential density drop off, where the wake morphology depends on the vertical depth of the transit within the disc. The inclination damping $\Delta i$ relative the pre-transit inclination behaves as a power law in $\sin(i)$ and the ambient Hill mass $m_\text{H,0}$ as $\Delta i/i \propto m_{\rm H,0}^{0.4} \sin(i)^{-2.7}$. The drag on the BH is dominated by the gravity of the wake for the majority of our inclination range until accretion effects become comparable at $\sin(i)\gtrsim30H_0/R_0$, where $H_0/R_0$ is the disc aspect ratio. At low inclinations ($\sin(i)\lesssim3H_0/R_0$) the wake morphology becomes more spherical, leading to a regime change in the inclination damping behaviour. Our results suggest that the inclination damping timescale is shorter than expected from only episodic Bondi-Hoyle-Lyttelton accretion events during each transit, implying inclined objects may be captured by the AGN disc earlier in its lifetime than previously thought. 
\end{abstract}

% Select between one and six entries from the list of approasssssssssaasved keywords.
% do not make up new ones.
\begin{keywords}
binaries: general -- transients: black hole mergers -- galaxies: nuclei -- Hydrodynamics -- Gravitational Waves
\end{keywords}

%%%%%%%%%%%%%%%%%%%%%%%%%%%%%%%%%%%%%%%%%%%%%%%%%%

%%%%%%%%%%%%%%%%% BODY OF PAPER %%%%%%%%%%%%%%%%%%

\section{Introduction}
\label{sec:intro}
It is expected that AGN discs will modify the angular momentum distribution of objects (stars and black holes) in the Nuclear Stellar Cluster (NSC) that cross through the highly dense accretion disc, within the lifetime of the AGN. Previous analytic \citep[][]{Syer1991,Mckernan2012,Bartos2017,Yang2019,Macleod2020,Fabj2020} and semi-analytic \citep[][]{Tagawa2020,Rowan2024_rates,Xue2025} studies have shown that objects that transit through the AGN disc can experience significant inclination damping due to dynamical gas drag and linear momentum accretion from the gas. As a result, an initially isotropic distribution will become flattened along the axis parallel to the angular momentum of the disc, with many objects initially on lower inclinations becoming fully embedded in the AGN disc. 

AGN could host a non-negligible fraction of compact object mergers \citep[e.g][]{Bartos2017,Yang2019_lett,Yang2019,Tagawa2020,Tagawa2020_spin,Tagawa+2021_eccentricity,McKernan2020_AGNmontecarlo,McKernan2020,Rowan2024_rates,Delfavero2024,Xue2025} detected by current \citep[e.g][]{LIGO2016,LIGO2019,LIGO2020a,LIGO2020b,LIGO2020e,LIGO2020c,LIGO2020d,abbott2022,abbott2022_fermi,KAGRA:2021duu, KAGRA:2021vkt} and future \citep[e.g][]{Hild2008,Amaro2017,Reitze2019,Adhikari2020} gravitational wave (GW) detectors. Mergers in the so called "AGN channel" are facilitated by a diverse array of phenomena including black hole binary formation through the 'gas-capture' mechanism \citep[][]{Li_Dempsey_Lai+2022,Rowan2022,Rowan2023,Whitehead2023,Whitehead2023_novae,Whitehead2025_adiabatic,Delaurentiis2023,Rozner2022} and Jacobi captures \citep[e.g][]{Trani2019,Trani2019_trip2,Boekholt2023}, BH binary-single encounters with \citep[e.g][]{Rowan2025,Wang2025} and without gas \citep[e.g][]{Leigh2018,Ginat2021,Fabj2024} and binary hardening through gas induced torques \citep[][]{Baruteau2011,Li2021,Li_2022_hot_discs,Li2022,Li_and_Lai_2022_windtunnel_II,Li_and_Lai_windtunnel_III_2023,Dempsey2022,Rowan2022,Vaccaro2024,Dittman2024,ONeill2024}. 

The literature suggests that many merging BHBs will have formed through the gas-capture mechanism and then brought to merger through a combination of gas torques and binary-single interactions. The number and physical parameter distribution (i.e mass, mass ratio etc) of merging binaries in the AGN channel is sensitive to several parameters, but primarily the achievable number density of objects embedded in the disc \citep[e.g][]{Rowan2024_rates} and the AGN lifetime $t_\text{AGN}$ \citep[][]{Delfavero2024,Xue2025}. The number density governs the frequency of object interactions and $t_\text{AGN}$ gives the timescale over which they will have gas to drive orbital migration through the disc and encounter each other \citep[][]{Secunda2019,Secunda2021}, assist in binary formation and dissipate their orbital energy thereafter. The embedded number density across the AGN disc will increase over time as more objects damp their inclination, until a balance is achieved between objects embedding themselves and the reduction due to mergers and the scattering of objects back out of the disc through scattering \citep[][]{Delfavero2024,Xue2025}. The anticipated number density depends intimately on how efficiently the inclination is damped. Typically in the literature, an inclined BH's orbital elements are evolved according to Bondi-Hoyle-Lyttelton (BHL) drag, where $t_\text{damp}$ is inversely proportional to the accreted mass per transit. However, whether the true drag on the BH can be accurately captured by BHL drag (which assumes the BH is moving with constant velocity through an infinite uniform medium) remains to be tested.

In this work we perform high resolution radiative adiabatic hydrodynamical simulations of BH's on inclined orbits crossing the AGN disc. We perform a total of 79 simulations covering inclinations in the range $i \in \left[2^\circ-15^\circ\right]$. This work is presented alongside a sibling paper \cite{Whitehead2025}, hereafter \citetalias{Whitehead2025}, which covers the low inclination regime $i \in \left[0^\circ-2^\circ\right]$. Together, these two papers simulate the orbital re-alignment of moderately to lowly inclined BHs, moving towards a more accurate model for the alignment time for inclined objects in AGN. The structure of this paper is as follows. We outline our computational methods and assumptions in Sec. \ref{sec:methods}. We present our results in Sec. \ref{sec:results}. The implications and limitations of our findings are discussed in Sec. \ref{sec:discussion}. We summarise our results in Sec. \ref{sec:conclusions}.

\section{Methods}
\label{sec:methods}  
The system is modelled using the Eulerian GRMHD code \texttt{Athena++} \citep{Stone_2020}, simulating a portion of the AGN disc in 3D as a shearing box \citep[e.g][]{Goldreich1978,Hawley1994,Hawley1995}. The natural length scale of the system is the Hill sphere of the binary
\begin{equation}
     r_{\rm H}=\,R_0\bigg(\frac{\mBH}{3\mSMBH}\bigg)^{1/3}\,,
     \label{rh2}
    \centering
\end{equation}
where $R_0$ is the radial position of the shearing box centre from the SMBH, $M_\bullet$ is the SMBH mass and $\mBH$ is the BH mass. 
\subsection{The shearing box}
  The shearing box lies in a non-inertial reference frame that co-rotates with the AGN disc at a fixed radius $R_0$ and angular frequency $\Omega_0=\sqrt{G\mSMBH/R_0^{3}}$. The Cartesian coordinate system of the shearing box $\{x,y,z\}$ can be translated to a position in the global AGN disc in cylindrical coordinates $\{R,\phi,z\}$ via
\begin{equation}
    \boldsymbol{r}=\Bigg(\begin{matrix}
        R \\ \phi \\ z
    \end{matrix}\Bigg)
    =\Bigg(\begin{matrix}
         R_0 + x  \\
        \Omega_0 t+y/R_0  \\
        z
    \end{matrix}\Bigg)\,.
    \label{eq:transformation}
\end{equation}
The gas and BH are subject to accelerations from the SMBH in the co-rotating frame according to
\begin{equation}
    \label{eq:a_smbh}
    \boldsymbol{a}_{\text{SMBH}}= 2 \boldsymbol{u} \times \Omega_0 \hat{z} + 2q\Omega_0^2 (x-x_\text{sh})\hat{x}-\Omega_0^{2} z\,\hat{z}\,,
\end{equation}
where $\boldsymbol{u}$ is the object's velocity, and $q=-\frac{d\ln\Omega}{d\ln R}=\frac{3}{2}$ is the velocity shearing term for a Keplerian disc and $x_\text{sh}$ is centre of the shearing frame, i.e the co-rotation radius. 

The boundary conditions at the $x$ and $z$ boundaries are set to outflow, i.e the flow properties are reset to their initial values at the start of the simulation. The inflow regions assume the initial gas velocities and densities at the start of the simulation; $\Sigma=\Sigma_{0}$ (see Sec. \ref{sec:AGNdisc}) and $\{u_\mathrm{x},u_\mathrm{y},u_\mathrm{z}\}=\{0,-q\Omega (x-x_\text{sh}),0\}$.  
\begin{equation}
    (y=y_\mathrm{max}):\,\,\begin{cases}
        \text{outflow} & x<x_\text{sh},\\
        \text{refill} & x>x_\text{sh},\\        
    \end{cases}
\end{equation}
\begin{equation}
    (y=y_\mathrm{min}):\,\,\begin{cases}
        \text{refill} & x<x_\text{sh},\\
        \text{inflow} & x>x_\text{sh}.\\        
    \end{cases}
\end{equation}
The dimensions of the shearing box are $\{\Delta x,\Delta y,\Delta z\}=\{17 \times 34 \times 13.5\}$ $H_{0}$, where $H_{0}$ is the disc scale height, see Sec. \ref{sec:AGNdisc}.
\subsection{Gas dynamics}
\subsubsection{Equations of motion}
\label{sec:gas_dynamics}
\texttt{Athena++} solves the fluid equations in Eulerian form through the extended Navier-Stokes equations: 
\begin{align}\label{eq:EOM}
    &\frac{\partial \rho}{\partial t} + \nabla \cdot \left(\rho \boldsymbol{u}\right) = 0\,,\\
    &\frac{\partial \left(\rho \boldsymbol{u}\right)}{\partial t} + \nabla \cdot \left(\rho \boldsymbol{u} \boldsymbol{u} + P \boldsymbol{I} + \boldsymbol{\Pi}\right) = \rho\left(\boldsymbol{a}_{\text{SMBH}} - \nabla \phi_\text{BH}\right)\,.\label{eq:EOM2}
\end{align}
Here, $\rho$, $\boldsymbol{u}$, $P$, $\Pi$, are the cell gas density, velocity, pressure and viscous stress tensor
\begin{equation}\label{eq:stress}
    \Pi_{ij} = \Sigma \nu \left(\frac{\partial u_i}{\partial x_j} + \frac{\partial u_j}{\partial x_i} - \frac{2}{3}\delta_{ij}\nabla \cdot \boldsymbol{u}\right)\,,
\end{equation}
For simplicity and to reduce computational expense, we ignore any magnetic and viscous effects, i.e $\nu=0$.

The remaining $\nabla\phi_{\rm BH}$ term is the acceleration from the stellar mass BH
\begin{equation}
    -\nabla \phi_\text{BH}(\boldsymbol{r}) = 
    \mBH \,g\left(\frac{\boldsymbol{r}-\boldsymbol{r}_\text{BH}}{h}\right)\,, 
    \label{eq:BH_grav}
\end{equation}
where $g(\boldsymbol{s})$ is the gas gravitational softening kernel \citep[e.g][]{Price_2007}
\begin{equation}
    g(\boldsymbol{s}) = -\frac{G}{h^2}\hat{\boldsymbol{s}}
    \begin{cases}
    \frac{32}{3}s - \frac{192}{5}s^3 + 32s^4 & 0 < s \le \frac{1}{2}\,, \\
    -\frac{1}{15s^2} + \frac{64}{3}s - 48s^2 + \frac{192}{5}s^3 - \frac{32}{3}s^4 & \frac{1}{2} < s \le 1\,, \\
    \frac{1}{s^2} & s > 1\,.
    \end{cases}
    \label{eq:softening}
\end{equation}
We set the softening length to a hundredth of the Hill sphere of a single BH, i.e $h=0.01r_{\rm H}$. The small softening is necessary so that we do not soften on a scale larger than the standoff distance of the bow shocks produced, as this would incorrectly model the pressure balance at the shock front.
\subsubsection{Equation of state}
We evolve the energy equation according to
\begin{equation}
    \frac{\partial E}{\partial t} + \nabla \cdot \left[\left(E+P\right)\boldsymbol{u} + \boldsymbol{\Pi} \cdot \boldsymbol{u}\right] = \rho \boldsymbol{u} \cdot \left(\boldsymbol{a}_\text{SMBH} - \nabla \phi_\text{BH}\right)\,.
    \centering
    \label{eq:EOS}
\end{equation}
Here $P$ is the pressure, $E$ is the fluid energy per unit volume, comprised of kinetic $K$ and internal $U$ components,
\begin{equation}
    E = K + U =  \frac{1}{2}\rho \boldsymbol{u} \cdot \boldsymbol{u} + U\,.
    \centering
\end{equation}
We utilise an adiabatic, radiative equation of state where the pressure and energy density have a gaseous and radiation contribution,
\begin{equation}
    P = P_\text{gas}+P_\text{rad}=\frac{k_\text{B}}{\mu_\text{p}m_\text{p}
    }\rho T + \frac{1}{3}a_\text{rad}T^4\,,
    \centering
\end{equation}
\begin{equation}
    U = U_\text{gas}+U_\text{rad}=\frac{3}{2}\frac{k_\text{B}}{\mu_\text{p}m_\text{p}
    }\rho T + a_\text{rad}T^4\,.
    \centering
\end{equation}
Here $k_\text{B}$, $\mu_\text{p}$, $m_{p}$, $a_\text{rad}$ are the Boltzmann constant, mean molecular weight, proton mass and radiation constant respectively. 
The adiabatic sound speed $c_\text{s}$ for the gas-radiation mixture can be calculated from the first adiabatic exponent $\Gamma$ \citep[e.g][]{Chandrasekhar_1939} 
\begin{equation}
    c_\text{s}^{2}=\Gamma\frac{P}{\rho}\,,
\end{equation}
\begin{equation}
    \Gamma = \frac{32-24\beta-3\beta^{2}}{24-21\beta}\,,
\end{equation}
\begin{equation}
        \beta=\frac{P_\text{gas}}{P_\text{gas}+P_\text{rad}}\,,
        \label{eq:beta}
\end{equation}
such that $\Gamma\rightarrow \frac{4}{3} (\frac{5}{3})$ for $\beta\rightarrow 0(1)$. 
To avoid the computational expense calculating the temperature through multiple inversions per cell per timestep (a significant computational burden) a lookup table (LUT) of thermodynamic values is instead passed to \texttt{Athena++} \citep[][]{Coleman_2020} from which the code can then interpolate the values, providing directly the required mappings of
\begin{equation}
    P = f_1(\rho,E)\,,
\end{equation}
\begin{equation}
    E = f_2(\rho,P)\,,
\end{equation}
\begin{equation}
    c_\text{s} = f_3(\rho,P)\,.
\end{equation}
As memory access time is independent of the size of an array, we can provide large (1000$\times$1000) LUTs which provide great accuracy without an increase in runtime. We do not expect the minor deviations from analytical values to have any significant effect on the hydrodynamics. 
\subsection{The black hole}
The black hole is represented by a point-like particle, which responds to the gravity of the gas (see Eq. \ref{eq:BH_grav}) and acceleration due to the SMBH (Eq. \ref{eq:a_smbh}). We also consider drag on the BH via the accretion of gas during the transit. The BH accretes momentum from the gas within a sink radius which we set equal to the softening radius, $r_\text{sink}=h$.  We compute the accelerations due to accretion by projecting the linear momentum of each gas cell within $r_\text{sink}$ onto the radial vector from the BH
\begin{equation}
    \boldsymbol{p}_\text{r}= \sum_{i}^{N_\text{acc}}(m_i (\boldsymbol{v}_i-\boldsymbol{v}_\text{BH})\cdot(\boldsymbol{r}_i-\boldsymbol{r}_\text{BH}))\frac{\boldsymbol{r}_i-\boldsymbol{r}_\text{BH}}{\|\boldsymbol{r}_i-\boldsymbol{r}_\text{BH}\|^{2}}\,,
\end{equation}
so as not to accrete any angular momentum from the gas \citep[e.g][]{Dempsey2020}. The acceleration from accretion is then calculated via
\begin{equation}
    \boldsymbol{a}_\text{acc} = \frac{f_\text{acc}}{\Delta t}\frac{\boldsymbol{p}_\text{r}}{\mBH}\,,
\end{equation}
where $\Delta t$ is the simulation timestep. The remaining factor $f_\text{acc}$ is a scaling factor which limits accretion beyond the Eddington limit. 
\begin{equation}
    f_\text{acc}=\,\,\begin{cases}
        1 & M_\text{acc}\leq M_\text{Edd},\\
        \frac{M_\text{Edd}}{M_\text{acc}} & M_\text{acc}>M_\text{Edd},\\        
    \end{cases}
\end{equation}
\begin{equation}
    M_\text{Edd} = \dot{M} _\text{Edd}\Delta t=\frac{L_\text{Edd}}{\xi c^{2}}\Delta t\,,
    \label{eq:M_edd}
\end{equation}
where $L_\text{Edd}$ is the Eddington luminosity
\begin{equation}
    L_\text{Edd} = \frac{4\pi G\mBH m_\text{p}c}{\sigma_\text{T}}\,.
\end{equation}
The remaining terms are the speed of light $c$, the proton mass $m_\text{p}$, the Thompson cross section $\sigma_\text{T}$ and the efficiency of accretion onto the BH $\xi$, which we take to be 0.1. Note that we do not actually evolve the mass of the BH or deplete gas within the cells of the simulation. While this may slightly modify the mass flux around the BH, this is a negligible consideration for our Eddington limited accretion rate.

As we consider high inclinations in the BH orbit ($i\sim2^{\circ}-15^{\circ}$) relative to the disc, the small angle approximation of the shearing box breaks down when considering the full orbit of the BH. However, we make the assumption that any significant impulse on the BH only takes place when the BH transitions through the thin AGN disc. With this, we can ignore the lateral motion of the BH we would expect (in the frame of the shearing box) and model the orbit of the BH purely through changes in its momentum when the BH is within the shearing box domain, where the BH velocity and position components can be reliably described under the transformation of Eq. \ref{eq:transformation}. 
The inclination of the BH after each transit can be measured directly via 
\begin{equation}
    i = \arcsin\bigg(\frac{z_\text{max}}{R_0}\bigg)
\end{equation}
where $z_\text{max}$ is the maximum height reached by the BH from the box midplane after each transit.
\subsection{The azimuthal headwind}
We initialise the simulation with the BH $\{x,y\}$ position in the centre of the box. The shearing frame is set such that the azimuthal velocity of the BH during its transit through the disc is zero in the reference frame of the box. To accomplish this, the shearing centre of the box is given by $x_\text{sh}=\Delta v_\text{y}/(q\Omega_0)$ where $\Delta v_\text{y}=v_\text{Kep}[1-\cos(i)]$, $v_\text{Kep}=\sqrt{G\mSMBH/R_0}$. This means the BH, which has initially zero $\hat{y}$ velocity component in the simulation frame, will experience a headwind in the $\hat{y}$ direction from the gas with velocity $\Delta v_\text{y}$ as it passes through the gas. 
\subsection{Mesh Refinement}
We apply an adaptive mesh refinement (AMR) procedure, allowing us to resolve the area around the BHs to a high degree, whilst minimising compute time and resources. The mesh closer to the location of the BH becomes more refined, down to a maximum refinement level. In all simulations shown here, we maintain a base resolution of 192x768x192 with a minimum refinement level of 6. For our box size, this gives a maximal and minimal cell size of $\delta_\mathrm{max}\simeq0.028\rh$ and $\delta_\mathrm{min}\simeq 0.00086\rh$, The softening/accretion length is resolved by $(h/\delta_\mathrm{min})^3\gtrsim1300
$ cells. The AMR scheme is centred on the BH and moves with it throughout the simulation, ensuring the flow around the BH is resolved at all times. The refinement algorithm begins refinement at a distance of 0.5$\rh$ from the BH and reaches maximal refinement at a distance of 0.2$\rh$. Outside of this radius, the resolution transitions to its base value.
\subsection{Initial conditions}
\subsubsection{The AGN disc}
\label{sec:AGNdisc}
We obtain a fiducial ambient surface density $\Sigma_{0}$ and sound speed $c_{\rm s,0}$ of the gas in the shearing box from AGN disc profiles generated using \texttt{pAGN} \citep{Gangardt2024} to ensure accurate AGN disc properties. In this paper, we consider a fiducial setup assuming an AGN disc with an Eddington fraction $L_\mathrm{\epsilon}=0.1$, radiative efficiency $\epsilon=0.1$ and hydrogen/helium fractions X/Y = 0.7/0.3. We set $\mSMBH=10^{7}\msun$ and the shearing box radius is set to $R_0=0.01$pc. For these parameters, this gives $\Sigma_\text{0}\simeq1.6\times10^{5}$kg$\,$m$^{-2}$, $c_{\rm s,0}\simeq11900$m$\,$s$^{-1}$ and disc thickness ratio $H_\text{0}/R_0\simeq0.0057$. In Sec. \ref{sec:param_sweep} we consider alternate ambient densities by modifying the fiducial value above. We perform four simulation suites, covering four disc densities $\Sigma = \{0.25,0.5,1,2\}\Sigma_\text{0}$.
\subsubsection{The BHs}
We maintain a fixed BH mass of $25\msun$ with initial inclinations in the range of $i \in \left[2^\circ-15^\circ\right]$, which are used to set their initial $z$ position via $z_\text{max,i} = R_0 \sin(i)$. The BH trajectories assume circular orbits with zero eccentricity. We limit ourselves to a maximum inclination of 15$^{\circ}$ as beyond this the Bondi-Hoyle-Lyttelton radius becomes comparable with our softening length.
\subsubsection{Repeated encounters}
We find that at an approximate inclination of $\sim3 ^\circ$ or below, the dynamics of a second encounter are affected non-negligibly by the perturbation to the disc from the first. This invalidates our assumption that we can capture the inclination damping by simulating individual transits.  Therefore we evolve these lower inclination systems continuously in time. This demands a significantly greater computational expense as the simulation must be integrated for longer and more notably the BH spends more time in the domain of the hydro mesh, which is highly refined around the BH. 

\section{Results}
\label{sec:results}
\subsection{Fiducial example ($i=5^{\circ)}$}
\begin{figure*}
    \centering
    \includegraphics[width=15cm]{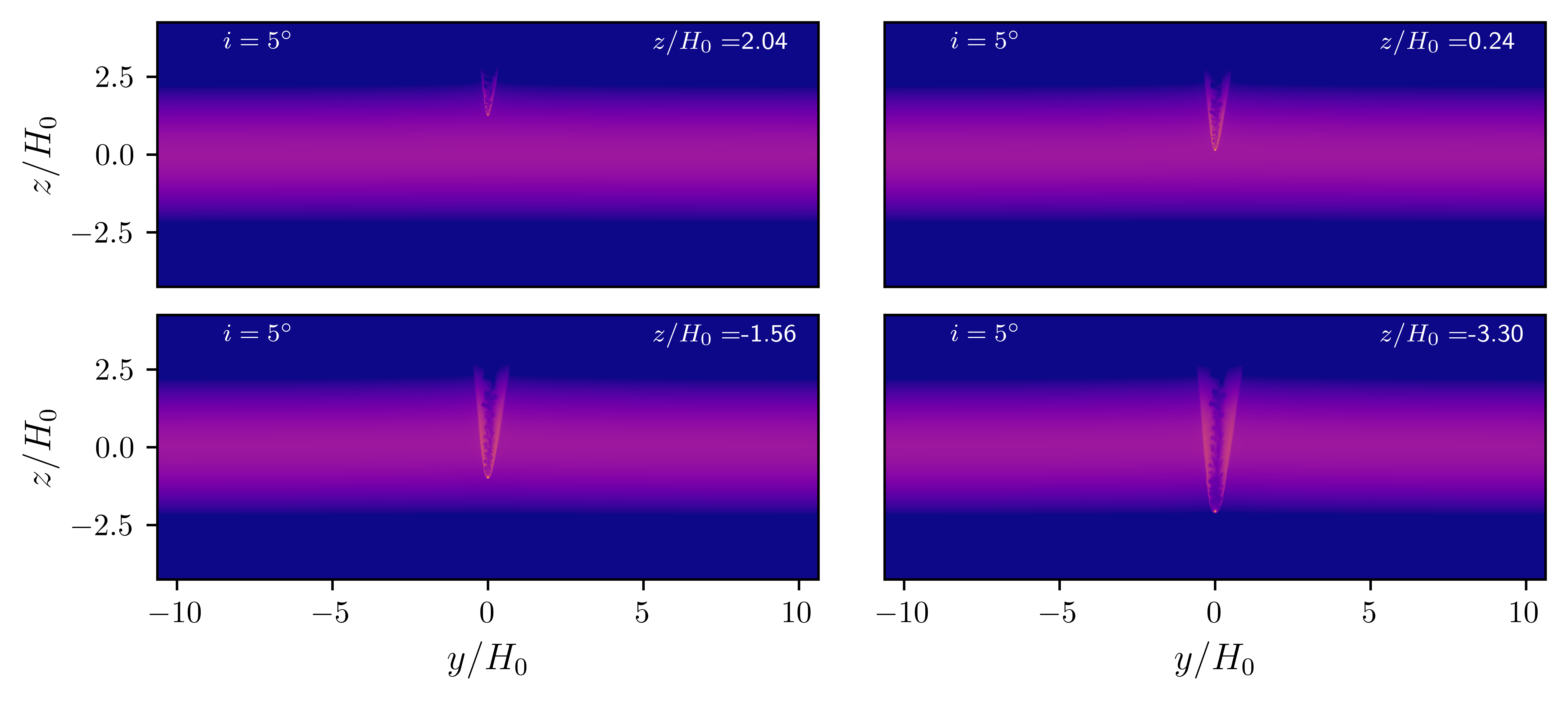}
    \includegraphics[width=16cm]{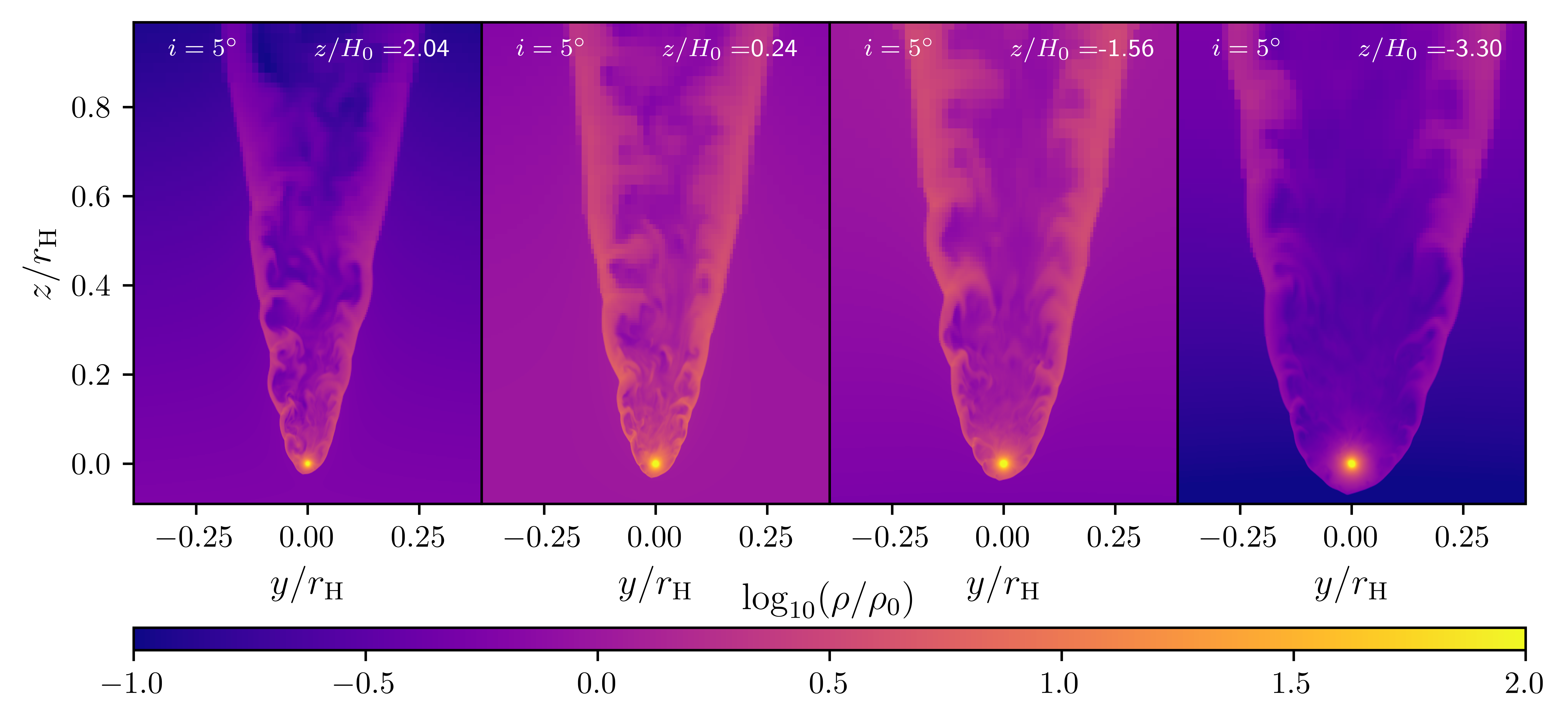}
    \caption{A timeseries of our fiducial model with $i=5^{\circ}$ and $\Sigma = \Sigma_0$, visualised through the 2D cross section in $\{y,z\}$ with $x$ centred on the BH position. \textit{Top panels:} the full simulation domain. \textit{Bottom panels:} close-up of the BH wake. We show the current vertical position $z$ in units of $H_0$ for each frame.}
    \label{fig:timeseries}
\end{figure*}

We first present an example transit from our fiducial run with an initial inclination of $5^{\circ}$. A visual timeline of the transit is shown through $\{y,z\}$ density slices of the simulation domain at $x=0$ in Figure \ref{fig:timeseries} (top panel) as well as zoom-ins of the wake morphology close to the BH (bottom panel). 
The high vertical velocity of the BH as it begins to penetrate the upper atmosphere of the disc ($z=2.5H_0$) leads to the formation of a strong bow shock that continues to propagate through the disc after the BH exits on the opposing side. The flow immediately behind the BH remains highly volatile with Kelvin-Helmholtz flow instabilities generated on the inner edges of the shock front that trails the BH. This is consistent with previous works that include radiation pressure in Bondi-Hoyle-Lyttelton (BHL) like flows \citep[e.g][]{Blondin1990}.

\begin{figure*}
    \centering
    \includegraphics[width=16cm]{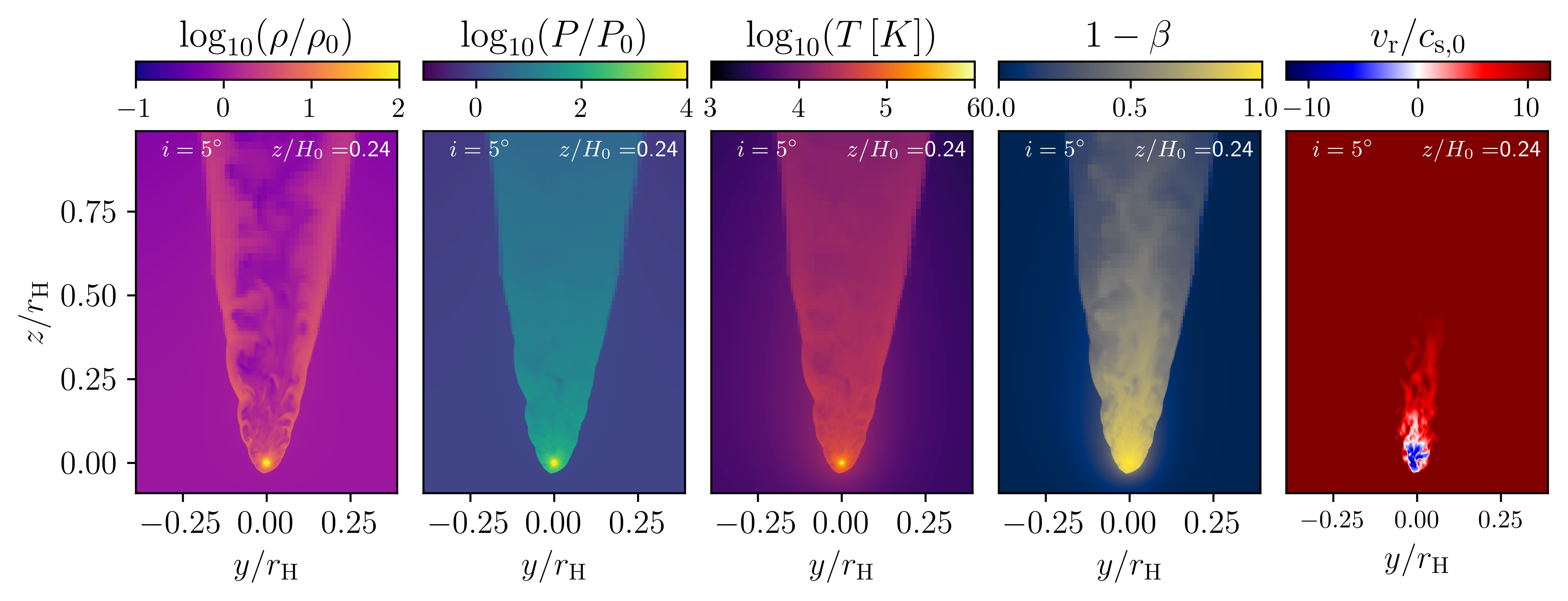}
    \caption{Snapshot of the gas properties in a $\{y,z\}$ cross section centred on the BH. \textit{Left to right:} the density relative to the initial midplane density $\rho_0$, the pressure relative to the midplane pressure $P$, the temperature, the pressure ratio $\beta$ (Eq. \ref{eq:beta}), the radial gas velocity relative to the BH.}
    \label{fig:wake}
\end{figure*}

We display additional physical properties of the wake at $z/H_0=0.21$ in Figure \ref{fig:wake}. These are the pressure, temperature, the ratio of gas pressure to total pressure $\beta$ and the radial velocity of the gas with respect to the BH $v_\text{r}$.
The pressure contrast across the shock front is approximately two to three orders of magnitude, with a similar contrast in temperature. We find the pressure is intensely radiation dominated ($1-\beta>0.9$) at the shock front ahead of the black hole and remains important ($1-\beta\gtrsim0.5$) in the trailing wake out to a distance of $\sim\rh/2$. This finding applies to all of simulations in this study. Therefore we encourage future studies to include the effects of radiation pressure, as we will see this region dominates the gravitational drag on the BH (Sec. \ref{sec:dissipation_sources}).

Looking at the last panel of Figure \ref{fig:wake}, we note the absence of a well defined BHL tail in the flow around the BH, despite our softening radius and cell size being $\sim 1/20 R_\text{BHL}$ and $\sim1/200 R_\text{BHL}$, where $R_\text{BHL}$ is the BHL radius given by 
\begin{equation}
    R_\text{BHL} = \frac{2G\mBH}{v_\text{rel}^2 + c_\text{s}^{2}}\,.
    \label{eq:RBHL}
\end{equation}
These size comparisons assume the BHL radius is at its minimum value over the course of the orbit (i.e where $v_\text{rel}$ is maximised), which occurs at the crossing of the midplane where $v_\text{rel}=2\sqrt{GM_\bullet/R_0}\sin(i/2)$. In principle, the BHL radius is larger further from the midplane as $\dot{z}$ is reduced. The chaotic flow close to the BH in our simulations prevents the formation of a steady and well defined accretion tail typical of a BHL flow, consistent with previous simulations of BHL flows which include radiation pressure \citep[e.g][]{Blondin1990,Edgar2004}. The BHL formalism also assumes the medium is uniform, where as in our case the BH passes through a density gradient spanning six orders of magnitude (midplane to background density) in a short timescale. \cite{Krumholz2006} report that BHL accretion can be easily exceeded when the medium is turbulent, suggesting that the flow may not be accurately described by BHL drag when the density is changing rapidly. 
\subsection{Dissipation sources}
\label{sec:dissipation_sources}
To assess the work done by the gas against the vertical motion of the BH, we define the "vertical" energy of the BH as 
\begin{equation}
    E_\mathrm{z} = \frac{1}{2} \dot{z}^{2} + \frac{1}{2}\Omega_0^2 z^{2}
    \label{eq:E_z}
\end{equation}
The vertical energy dissipation rates for gravitational and accretion drag are then given by
\begin{equation}
    \varepsilon_\text{gas,z}= a_\text{gas,z}v_\text{z}\,,
    \label{eq:gas_drag}
\end{equation}
\begin{equation}
    \varepsilon_\text{acc,z}= a_\text{acc,z}v_\text{z}\,,
    \label{eq:acc_drag}
\end{equation}
where $a_\text{gas,z}$ and $a_\text{acc,z}$ are the vertical accelerations on the BH due to gas and accretion respectively and $v_\text{z}$ is the BH's vertical velocity. One can extract the instantaneous inclination by computing the anticipated height the BH will reach according to its current $E_\text{z}$ within the SMBH potential.
\begin{align}
    i &= \arcsin\bigg(\frac{z_\text{max}}{R_0}\bigg)
    %\nonumber\\   &
    =\arcsin\Bigg(\sqrt{\frac{2E_\text{z}}{\Omega_0^2}}\frac{1}{R_0}\Bigg)
    %\nonumber\\   &=
    =\arcsin\Bigg(\sqrt{\frac{\Omega_0^2z^2 + \dot{z}^{2}}{\Omega_0^2R_0^2}}\,\Bigg)
\end{align}
Note that both $E_\text{z}$ and the instantaneous $i$ here both ignore the gravitation to the gas itself and so $E_\text{z}$ and $i$ will increase slightly as the BH approaches the midplane and decrease as the BH moves away from it. In the absence of accretion and perturbations to the initial symmetries of the 3D density structure of the disc in $\{x,y,z\}$ about the origin of the simulation, the values of $E_\text{z}$ and $i$ will be symmetric about $z=0$ and their value at each $|z|$ will be conserved.

\begin{figure}
    \centering
    \includegraphics[width=8cm]{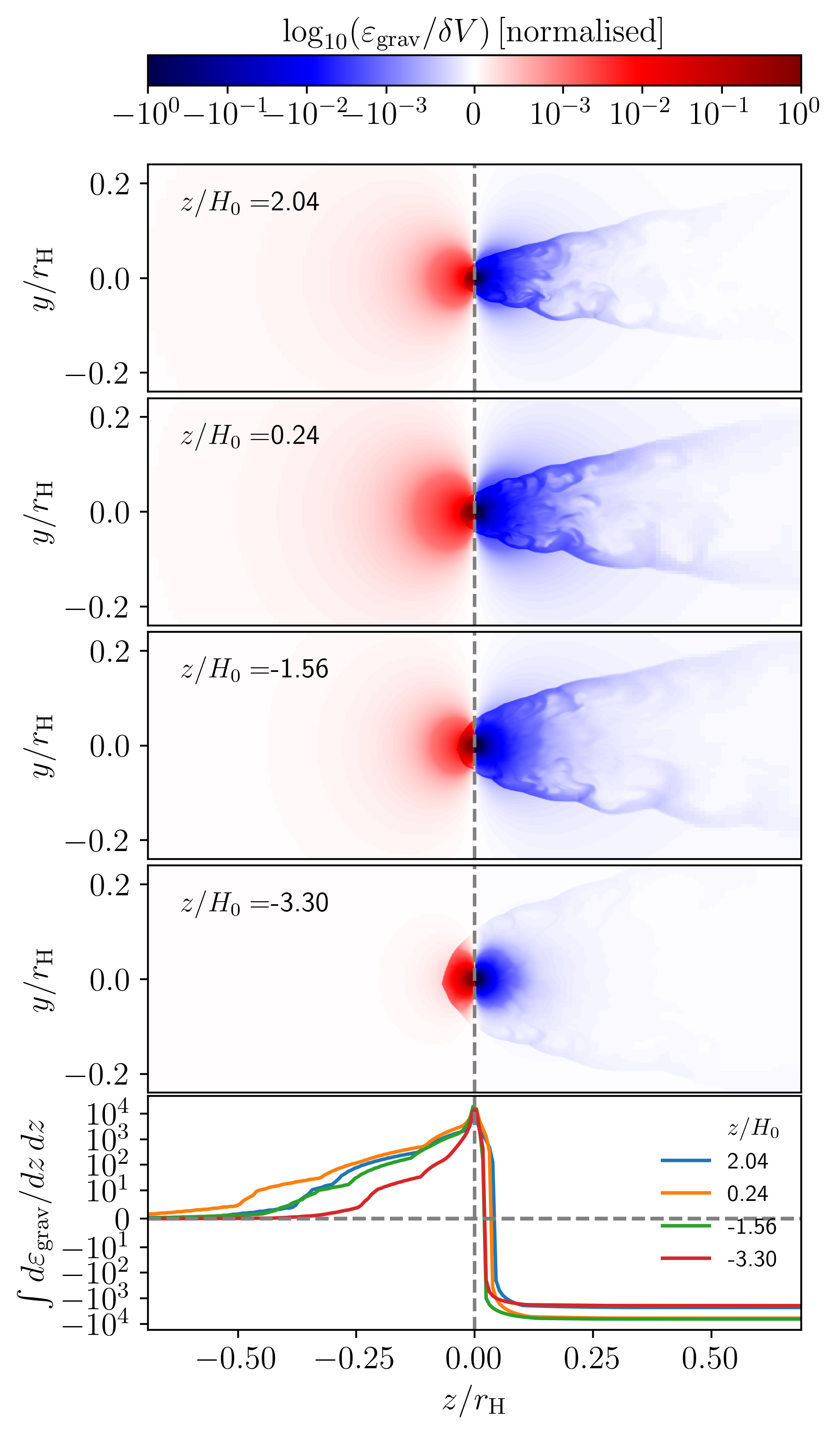}
    \caption{\textit{Top four panels:} cross sections of the contribution to $\varepsilon_\text{grav}$ per unit volume $\delta V$ visualised in the plane of $\{z,y\}$. \textit{Bottom:} the cumulative value of $\varepsilon_\text{grav}$ when summing over the contribution from cells in the $z$ direction contained within a cuboid of dimensions $|x/\rh| \leq 0.25$, $|y/\rh| \leq 0.25$ and $|z/\rh| \leq 0.7$ (the dimensions of the top panels).   }
    \label{fig:grav_wake}
\end{figure}

In figure \ref{fig:grav_wake} we show a density map of the contribution to $\varepsilon_\text{grav}$ from the BH wake as well as the cumulative value of $\varepsilon_\text{grav}$ when integrated along the wake.
Paying attention to the bottom panel, the high density of the shock front ahead of the BH does positive work on the BH $(\varepsilon > 0)$ while the tail does negative work. The magnitude of the BH's inclination change due to the gas gravity depends on the size of the imbalance between these two competing effects. The contribution from ahead of the BH (upstream) increases as the BH approaches the midplane and then decreases after the crossing. While there is a spike at the shock front, there is a non-negligible upstream contribution to $\varepsilon_\text{grav}$ from the ambient medium while the BH is still approaching the midplane, compared with the downstream contribution which is largely dominated by contributions close behind the BH throughout.
\subsection{Inclination damping as a function of inclination and disc density}
\label{sec:param_sweep}
\begin{figure}
    \centering
    \includegraphics[width=7cm]{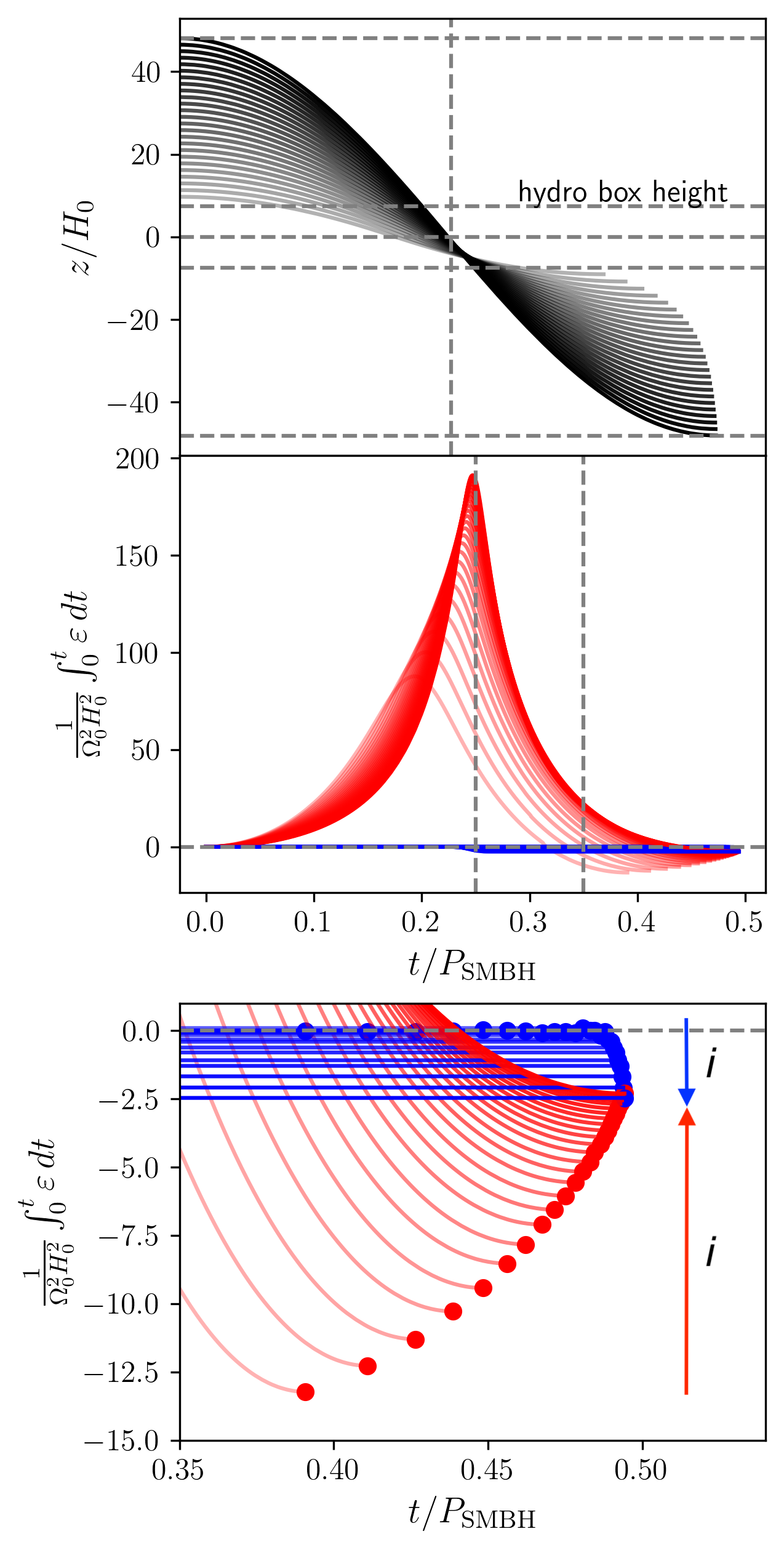}
    \caption{The black hole position and cumulative work done by gravity and accretion as a function of time, where $P_\text{SMBH}=2\pi/\Omega_0$. \textit{Top:} the BH $z$ position. \textit{Middle:} the cumulative work done on the BH by the gas due to gravitation (red) and accretion (blue). \textit{Bottom:} a zoom in of the cumulative work done as the BH completes its transit, reaching its new $z_\mathrm{max}$. The blue and red arrows indicate how the net work done by each dissipation mechanism evolves with increasing inclination.}
    \label{fig:inc_dep}
\end{figure}

We now consider transiting BHs with varying inclinations. In Figure \ref{fig:inc_dep} we show the $z$ position, cumulative gravitational work done $\varepsilon_\text{grav}$ (red) and cumulative accretion work done $\varepsilon_\text{acc}$ (blue) as a function of time.
The gravitation of the BH to the gas plane leads to an initial positive spike in the cumulative $\varepsilon_\text{grav}$. The gravitation with the wake (see Figure \ref{fig:grav_wake}) breaks the symmetry of $\varepsilon_\text{grav}$ about the midplane, leading to a net negative work done on the BH, reducing its inclination. We find cumulatively $\varepsilon_\text{grav}$ decreases as $i$ increases towards zero. The opposite is true in the case of accretion, as the accretion becomes less spherically symmetric. As low inclination BHs spend more time in the disc and dynamical gas drag is more efficient, we observe a smaller and earlier peak in $\varepsilon_\text{grav}$ as strong dissipation occurs earlier in the transit. At low $i$, $\varepsilon_\text{acc}$ is negligible compared to $\varepsilon_\text{grav}$ but increases to comparable values at $i\sim15^{\circ}$, see the bottom panel of Figure \ref{fig:inc_dep}. 

\begin{figure}
    \centering
    \includegraphics[width=8cm]{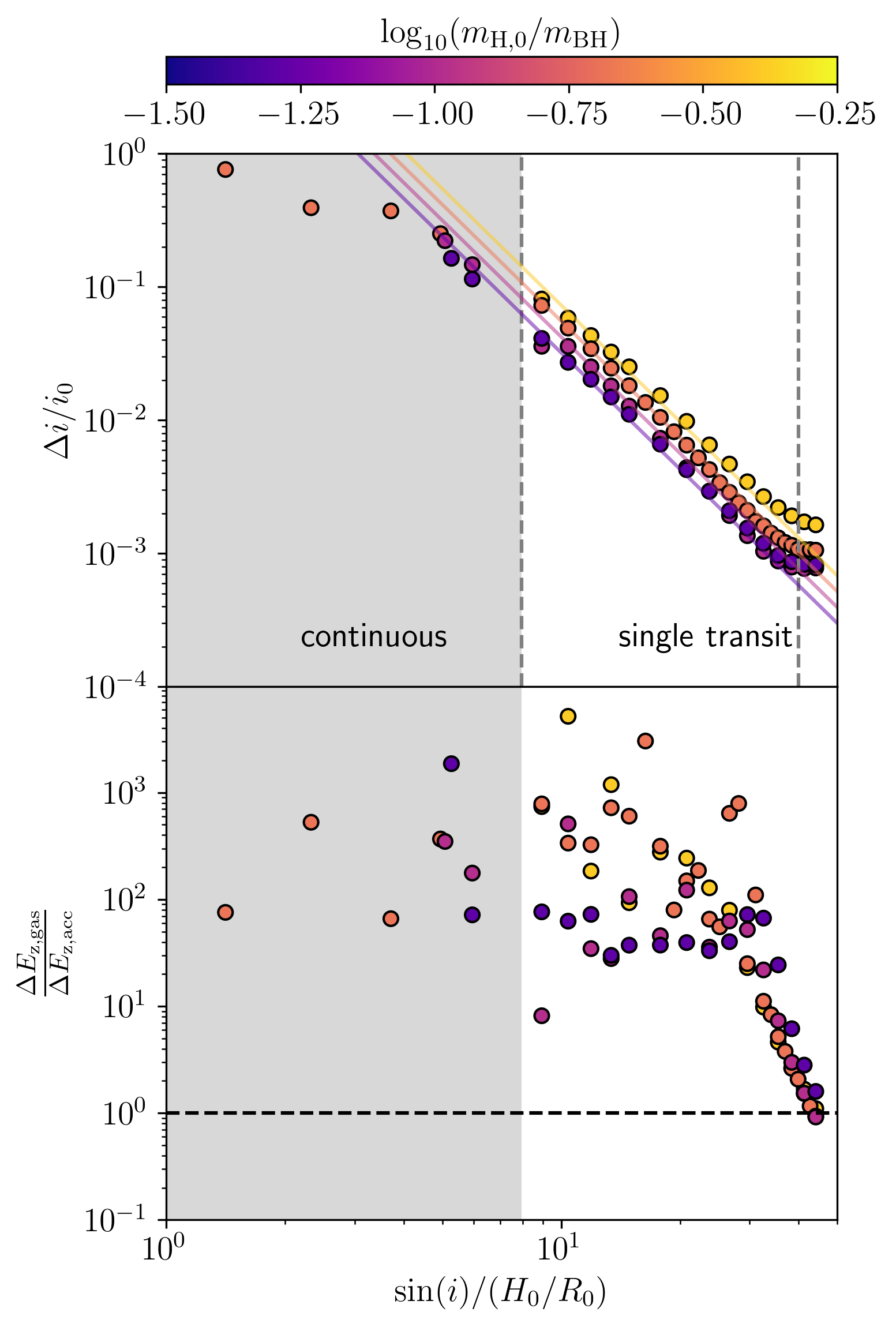}
    \caption{\textit{Top:} the relative change in inclination $\Delta i/i$ for each transit as a function of the inclination $i$. The solid lines represent the power law function of Eq. \eqref{eq:p_law_fit} using our best fit parameters. The vertical dashed lines highlight the inclination range for the data points used in the fit. \textit{Bottom:} the net work done on the BH by drag in the vertical direction due to gas (Eq. \ref{eq:gas_drag}) vs accretion (Eq. \ref{eq:acc_drag}).}
\label{fig:density_dependence}
\end{figure}

The negative work done in the vertical direction leads to a reduction in the BH inclination. In Figure \ref{fig:density_dependence} we show the net fractional change in inclination and the ratio of the net change in $E_\text{z}$ due to gas and accretion integrated over the transit as a function of inclination. The data is colour coded according to the relative initial Hill mass of the BH in the simulation 
\begin{equation}
    m_\mathrm{H,0}/\mBH=2\pi\rh^2\Sigma/\mBH\,,
    \label{eq:Hill_mass}
\end{equation}
as this has been shown to be the most influential parameter for gas driven dissipation for satellite interactions in accretion discs \citep[e.g][]{Whitehead2023,Whitehead2025_adiabatic}.
Driven by the decreasing gravitational drag, the relative inclination change decreases as a function of $i$. At the low inclination end $\sin(i)\lesssim 3 H_0/R_0$ we find the inclination change per orbit starts to become less steeply dependent on $i$ (see Sec \ref{sec:low_i_lim}). At our highest inclinations $\sin(i)\gtrsim30 H_0/R_0$ the inclination damping appears to undergo a regime change, with a more shallow dependence on $i$. However, we caution that at these inclinations $R_\text{BHL}$ approaches the softening radius when the BH is at the midplane and $v_\text{rel}$ is maximised. Further simulations which push to even higher resolutions are necessary to distinguish if this is a physical or numerical effect. 

We fit a power law to the data in log space with the form
\begin{equation}
        \log_{10}\left(|\Delta i| / i\right)_\text{high} = a_1 \log_{10}\left(\tilde{m}\right) +b_1\log_{10}(\tilde{i})+ c_1\,,
    \label{eq:p_law_fit}
\end{equation}
where $\tilde{i}=\sin(i)R_0/H_0$ and $\tilde{m}=m_\text{H,0}/\mBH$.
The fit parameters are shown in Table \ref{tab:line_fit}.

\begin{table}
    \centering
    \begin{tabular}{ c c c c}
    \hline
     Parameter $X$     & $\mu_X$   & $\sigma_X$ & RMS $\delta$ \\ \hline \hline
      $a_1$             & 0.39      & 0.03 & 0.07\\
      $b_1$             & -2.73     & 0.04 & \\
      $c_1$             & 1.70      & 0.06 & \\ 
    \hline
    \end{tabular}
    \caption{Best fit parameters $\mu_\text{X}$ and their, 1-$\sigma$ uncertainties $\sigma_\text{X}$ and fit root-mean-square error for the fit given by Eq. \ref{eq:p_law_fit}.}
    \label{tab:line_fit}
\end{table}

%\begin{figure*}
%    \centering
%    \includegraphics[width=16cm]{figures/shock_resolution_big.pdf}
%    \includegraphics[width=16cm]{figures/shock_resolution_small.pdf}
%    \caption{Caption}
%    \label{fig:enter-label}
%\end{figure*}
\subsection{Comparison to analytical drag models}
\subsubsection{Drag models}
Understanding how quickly compact objects from the surrounding spherical-like NSC can embed in the AGN disc will affect the rate of compact object mergers by governing the density of embedded objects over time. The thin geometry of the disc $H/R\lesssim 0.01$ implies the vast majority of objects (assuming a spherically symmetric distribution) will exist outside the AGN disc when a galactic nucleus enters an active phase. Therefore even if a small fraction of objects can embed themselves within $t_\text{AGN}$, this could heavily alter the anticipated number of mergers. The majority of studies to date utilise analytical approaches to estimate the alignment time. The most popular approach is to treat each disc passage as an instantaneous impulse driven by accretion \citep[e.g][]{Yang2019,Fabj2020,Nasim23,Wang24,spieksma2025,Xue2025}, where the momentum of the BH is updated after crossing.
Alternatively, as in \cite{Bartos2017,Generozov23, Rowan2024_rates} for example, a simpler characteristic timescale can be determined by assuming the fractional change in vertical velocity $v_\text{z}$ scales with the fractional change in mass $\Delta m_\text{cross}$ per crossing such that
\begin{equation}
    t_\text{align} \sim \frac{\pi v_\text{z}}{\Delta v_\text{z}\Omega_0}\sim\frac{\pi \mBH}{\Delta m_\text{cross}\Omega_0}\label{eq:simple_t_align}
\end{equation}
The assumption is then the BH accretes mass in a cylinder of radius $R_\text{BHL}$ with $v_\text{rel}$ taken to be the relative velocity of the BH to the gas at the midplane $v_\text{mid}$. Folding this all in gives the alignment time as \citep[e.g][]{Rowan2024_rates}
\begin{align}
t_\mathrm{align}&=\frac{t_\mathrm{orb}}{2}\frac{\cos(i/2)( v_\text{mid}^{2}+c_s^{2})^{2}}{4G^{2}\mBH\pi\Sigma}\,,\label{eq:t_align_full}\\
v_\text{mid} &= 2\sqrt{\frac{GM_\bullet}{R_0}}\sin\bigg(\frac{i}{2}\bigg)\,, \label{eq:v_mid}
\end{align}
We give the full derivation and set of assumptions for this expression in Appendix \ref{app:t_align} for completeness.

Alternatively, the drag may be modelled by dynamically integrating the orbit with the drag expressions for BHL drag, Ostriker dynamical friction \citep[e.g][]{Ostriker1999} $a_\text{GDF}$, or some combination of both. The acceleration on an object moving with through a uniform medium of density $\rho$, constant isothermal sound speed $c_\mathrm{s}$ with relative velocity to the gas $v_\text{rel}$ under BHL drag is given by 
\begin{equation}
    \boldsymbol{a}_{\mathrm{BHL}} = -\dot{M}_\text{BHL}\boldsymbol{v}_\text{rel},
\end{equation}
where the accretion rate is given by
\begin{align}
    \dot{M}_\text{BHL}&=\pi r_c^2\rho v_\text{rel}\,,\label{eq:BHL_accretion} \\
    r_c &= \min(R_\text{BHL},\rh)\,,
\end{align}
such that in the high inclination limit where $R_\mathrm{{BHL}}<<r_\text{H}$, we have
\begin{equation}
    \boldsymbol{a}_{\mathrm{BHL}} = -\frac{4\pi G^2 \mBH \rho(z)}{(v_\mathrm{rel}^2+c_\text{s}^2)^{3/2}}\boldsymbol{v}_\mathrm{rel}\,,
    \label{eq:BHL_drag} 
\end{equation}
which is the case for the majority of our parameter space. For dynamical friction, the acceleration is given by
\begin{equation}
    \boldsymbol{a}_{\mathrm{GDF}} = -\frac{4\pi G^2 \mBH \rho(z)}{v_\mathrm{rel}^3}\boldsymbol{v}_\mathrm{rel} f\left(\mathcal{M}\right),
    \label{eq:GDF_drag}
\end{equation}
where $\mathcal{M} = v_\mathrm{rel} / c_s$ is the Mach number and $f(x)$ is expressed as 
\begin{equation}
    f(x) = \begin{cases}
        \frac{1}{2}\ln\left(\frac{1+x}{1-x}\right) - x & x < 1 - x_m, \\
        \frac{1}{2}\ln\left(\frac{1+x}{x_m}\right) + \frac{\left(x-x_m\right)^2-1}{4x_m} & 1-x_m \leq x < 1 + x_m, \\
        \frac{1}{2}\ln\left(x^2-1\right)+\ln\Lambda & x \geq 1 + x_m,
    \end{cases}
\end{equation}
where in this work we adopt $\ln \Lambda = -\ln x_m = 3.1$ following \citet{Chapon_2013}. We use eqs. \eqref{eq:BHL_drag} and \eqref{eq:GDF_drag} to calculate a lower bound on the alignment time by
evolving the vertical motion of a BH starting from $15^\circ$ and
ignoring the azimuthal and radial contributions to $v_\mathrm{rel}$.  With this assumption, we reduce the problem to a damped harmonic oscillator in $z$ with the accelerations given by
\begin{equation}
    a_\text{z,damp}=-\Omega_0^2z + a_\text{drag}\,,
    \label{eq:oscillator}
\end{equation}
where $a_\text{drag}$ is the selected drag formula of \eqref{eq:BHL_drag} or \eqref{eq:GDF_drag}. We note that much like in our simulations the general assumptions of BHL and Ostriker drag are not met (i.e $v_\text{rel}$, the local $\rho$ are not fixed in time and $\rho(z)$ is not uniform), but we nevertheless test the applicability of these expressions to our system.

\subsubsection{Alignment time comparisons}
\label{sec:alignment_times}
\begin{figure}
    \centering
    \includegraphics[width=8cm]{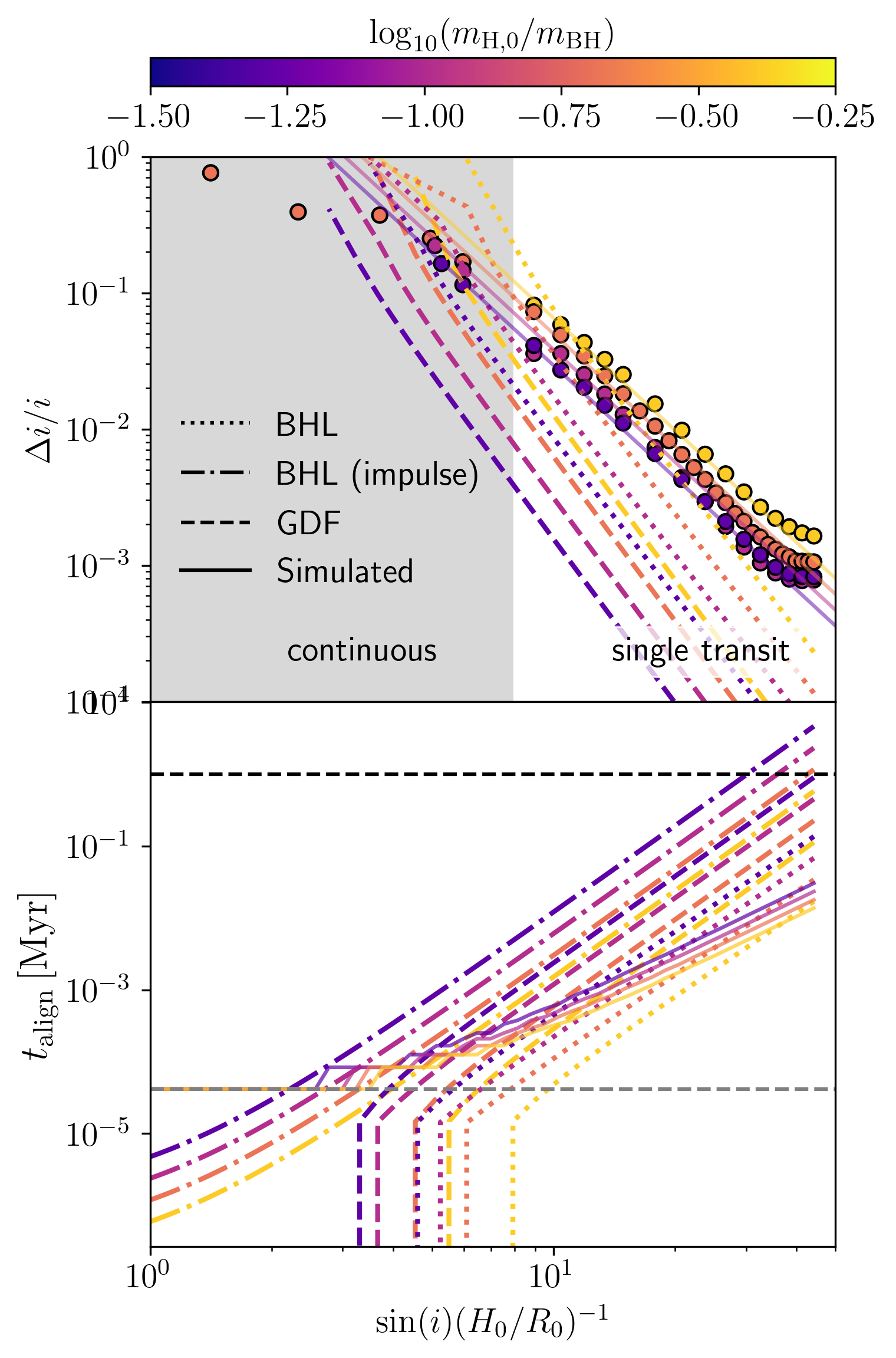}
    \caption{\textit{Top:} Comparison of the the relative inclination change $\Delta i/i$ as a function of $i$ between our simulations, the predictions of BHL (Eq. \ref{eq:BHL_drag}) and gas dynamical friction (Eq. \ref{eq:GDF_drag}). \textit{Bottom:} the anticipated alignment time as a function of inclination from our simulations alongside the predictions from BHL drag (from both the impulse approximation \eqref{eq:t_align_full} and damped oscillator \eqref{eq:BHL_drag}), and gas dynamical friction.}
    \label{fig:t_aligns}
\end{figure}

\begin{figure}
    \centering
    \includegraphics[width=8cm]{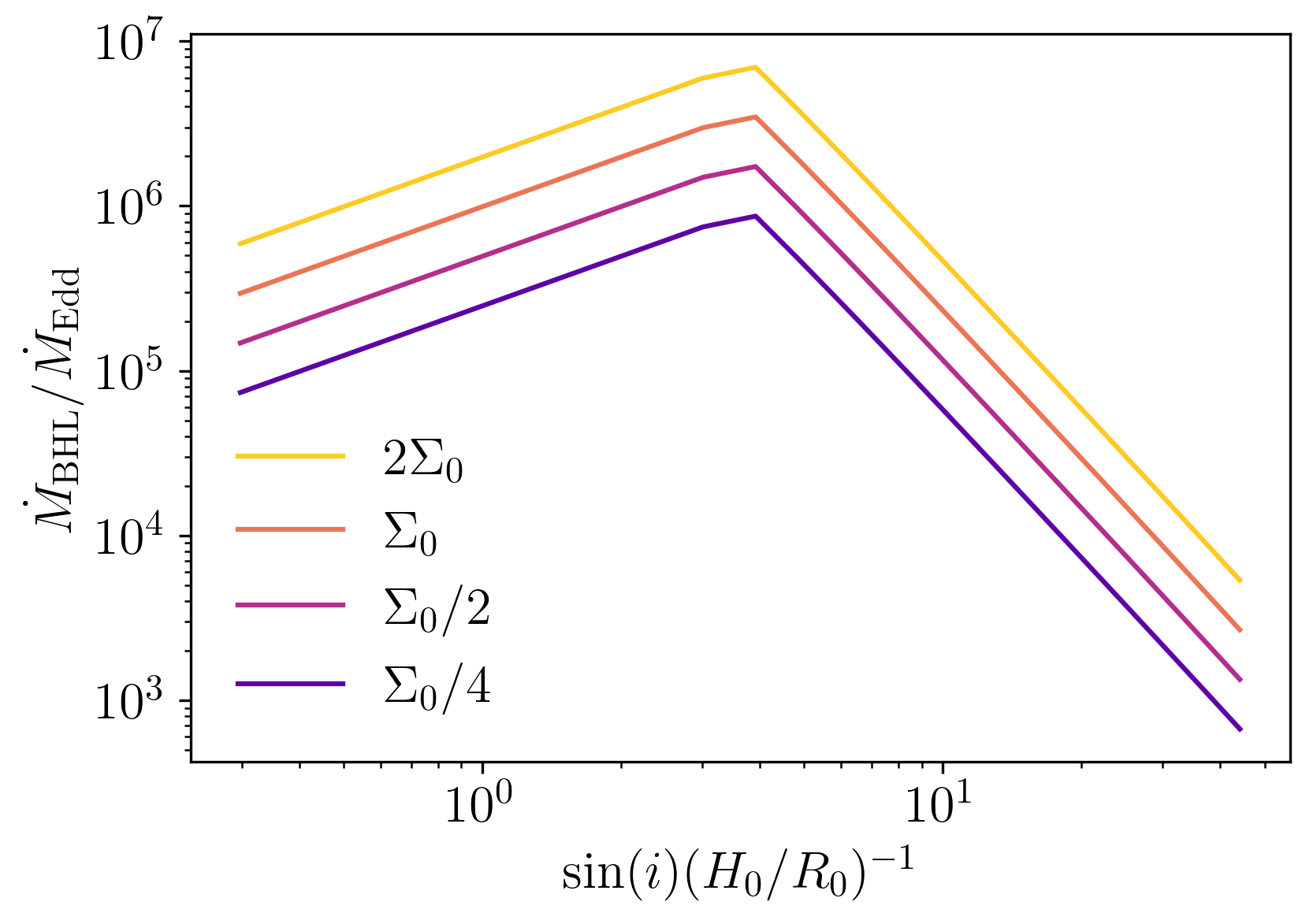}
    \caption{The ratio of the accretion rates expected from the BHL formalism (Eq. \ref{eq:BHL_accretion}) vs the Eddington accretion rate for our four disc densities.}
    \label{fig:BHL_to_Edd}
\end{figure}

In Figure \ref{fig:t_aligns} we compare the predicted alignment times inferred by our fit to the slope of $\Delta i/i$ with the alignment times predicted by the BHL impulse approximation and our simplified damped harmonic oscillators assuming Eqs. \eqref{eq:BHL_drag} and \eqref{eq:GDF_drag}. We calculate an alignment time from Eq. \eqref{eq:p_law_fit} by determining the number of crossings required to evolve from $i$ to $i=\arcsin(H_0/R_0)$, i.e when the BH's vertical motion no longer exceeds the scale height. We also compare the inclination change per crossing from the analytic BHL and Ostriker drag formulae with our simulation data.
Our results show that simply evolving $z(t)$ with Ostriker drag leads to a shorter alignment time compared to the other analytical models. The single orbit impulse approximation consistently predicts the longest alignment timescale. This is unsurprising since the impulse model makes three conservative assumptions: i) the rate of inclination change per orbit does not evolve as the inclination damps, ii) the momentum change is due to accretion alone and iii) $v_\text{rel}$ is fixed at it's maximal (midplane) value, minimising $R_\text{BHL}$. The simulations indicate dynamical gas drag can remain significant compared to BHL accretion for our inclination range. The alignment time predicted by our simulations lies between the predictions of the impulse and live BHL drag models. Due to the differing slope in $\Delta i/i$ in our simulations, the alignment timescale is in better agreement with the impulse approximation at low inclinations and the live GDF drag at higher inclinations. We posit that the smaller $t_\text{align}$ at low $i$ predicted from \eqref{eq:BHL_drag} results from hyper-accretion onto the BH. We show the ratio of $\dot{M}_\text{BHL}/\dot{M}_\text{Edd}$ according to Eqs. \eqref{eq:M_edd} and \eqref{eq:BHL_accretion} in Figure \ref{fig:BHL_to_Edd}.
The accretion predicted by BHL ranges from $10^{7}$ to $10^{3}$ $\dot{M}_\text{Edd}$ for $i=2^\circ$ to $i=15^\circ$. Given this discrepancy, it is natural that the simulated alignment time becomes longer than the BHL expectation as the BH becomes more aligned with the disc. The reduction in the accretion rate at low inclinations results from the accretion cross section transitioning from the BHL radius to the Hill radius.

The validity of such hyper-Eddington accretion rates, particularly at lower inclinations, is uncertain as radiation pressure is neglected in the construction of Eq. \eqref{eq:BHL_drag}. For example \cite{Kley1995} report significantly reduced drag compared to the BHL expectation when the wake is radiation dominated and optically thick (as given in this work). If accretion driven radiative feedback is significant enough, the acceleration can change sign and become positive \citep[e.g][]{Canto2013,Li2020_BHL}. Furthermore, \cite{Zanotti2011} performed general relativistic radiative simulations of an object moving through a radiation dominated gas medium, finding BHL to overestimate the accretion rate and drag by at least two orders of magnitude, yet the accretion is still super-Eddington with Eddington fractions between 1-7. We note that their ambient densities are around 5 orders of magnitude lower than our midplane density. Whether we should expect a similar accretion limit in our problem is therefore more uncertain.

To quickly summarise these comparisons, we find the alignment time is in reasonable agreement with BHL drag at low inclinations, but diverges at higher inclinations ($\sin(i)\gtrsim20H_0/R_0$, agreeing more with GDF models.
\begin{figure*}
    \centering
    \includegraphics[width=17cm]{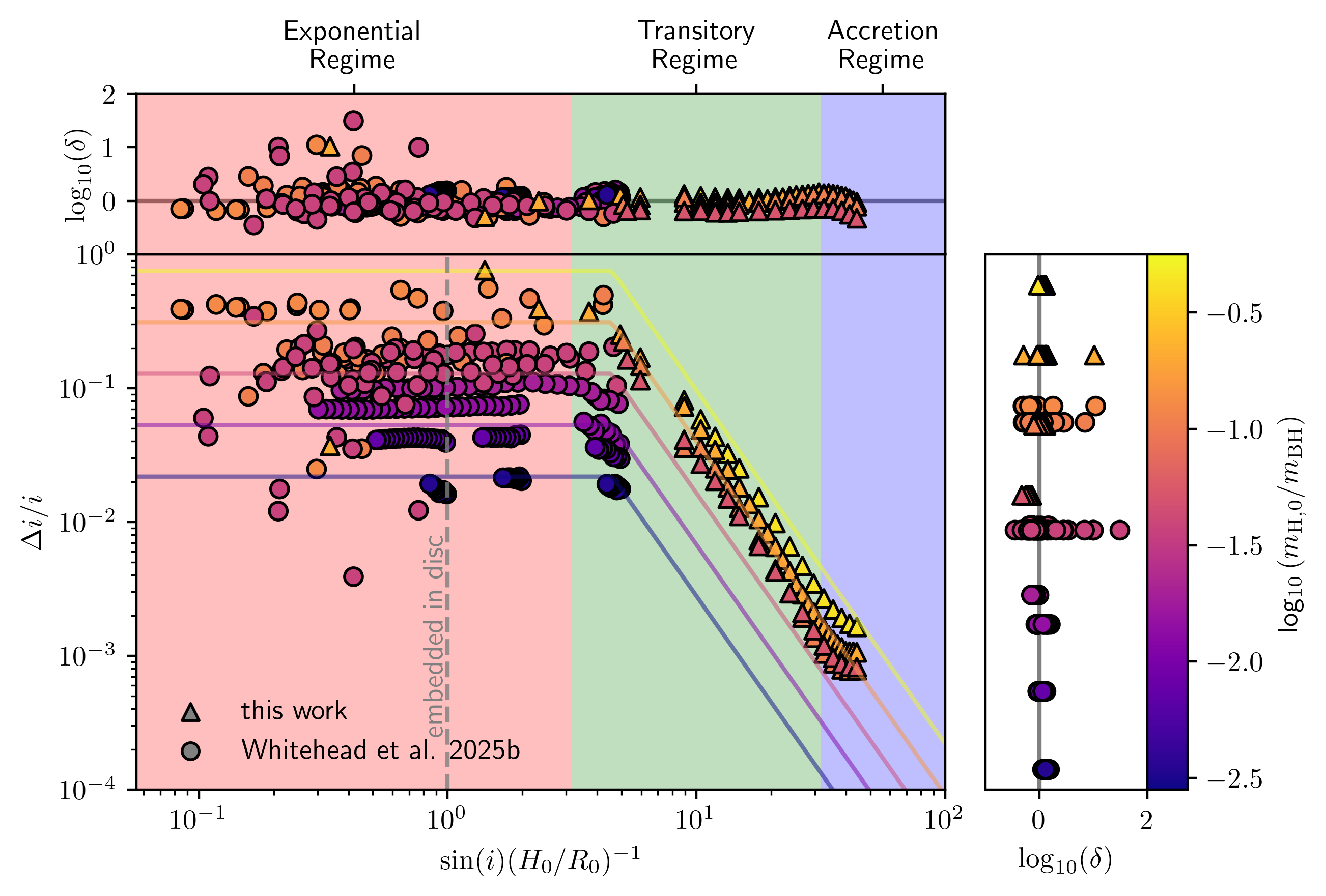}
    \caption{The combined data of this work and sibling paper \citetalias{Whitehead2025}, showing the relative inclination change $\Delta i/i$ per transit as a function of the pre-transit inclination. The markers are colour coded by the ambient Hill mass $m_\text{H,0}/\mBH$. We overlay our phenomenological model (Eq. \ref{eq:combined_fit}) for five values of $m_\text{H,0}/\mBH$ and highlight the three damping regimes discussed in Sec. \ref{sec:regimes}. We also show the residuals $\delta$ across both parameters in Eq. \eqref{eq:combined_fit}.} 
    \label{fig:di_vs_i_all}
\end{figure*}
\subsection{Inclination damping from high to low $i$}
\subsubsection{Inclination evolution in the low $i$ limit}
\label{sec:low_i_lim}
The sibling paper \citetalias{Whitehead2025}, presented simultaneously with this work, examines the inclination evolution continuously from lower inclinations of $\sin(i) \leq 3H_0/R_0$ until the BH becomes fully embedded in the AGN disc. There are several differences in the simulations of \citetalias{Whitehead2025} compared to this work, most notably \citetalias{Whitehead2025} include viscosity and neglected radiation pressure, accretion effects, the horizontal (azimuthal) velocity of the BH relative to the gas\footnote{This velocity becomes vanishingly small at such low inclinations.} and utilise a slightly larger softening and lower resolution. The initial conditions of \citetalias{Whitehead2025} were generated by changing $R_0$ and determining $\Sigma_0$ and $H_0$ self consistently from an AGN disc model constructed by \texttt{pagn} \citep[][]{Gangardt2024}, whereas this work scales $\Sigma_0$ while arbitrarily fixing all other parameters. Crucially this means that in \citetalias{Whitehead2025}, changes to the ambient Hill mass are accompanied by changes in other quantities e.g. radial position in the disc, sound speed and scale height. The evolution of the BH trajectories are performed continuously from their initial inclination until the termination of each simulation, covering many SMBH periods. Together, \citetalias{Whitehead2025} and this study cover a wide range of initial inclinations, spanning $i \in \left[0.1^\circ, 15^\circ\right]$. 

\citetalias{Whitehead2025} find the inclination damping in the low $i$ limit follows an exponential decay, with the damping modelled by a power law in $\tilde{m}$
\begin{equation}
    \log_{10}\left(|\Delta i| / i\right)_\text{low} = a_2 \log_{10}(\tilde{m}) + c_2\,,
    \label{eq:henry_fit}
\end{equation}
with $a_2=0.69\pm0.03$ and $c_2 = -0.09\pm0.06$. The strength of damping in the low inclination regime is independent of $i$, markedly different from the high $i$ regime, and resulting in an exponential decay in inclination over time. The Hill mass dependence $a_2=0.69$ is stronger than in the high $i$ regime $a_1=0.39$. We suspect this may result from the differing thermodynamics in our simulations here, where the wakes are acutely radiation pressure dominated. Alternatively, the Hill mass $m_\mathrm{H}$ may not provide the same scaling as the morphology of the perturbation to the gas by the BH is different between the inclination regimes considered by each work (see Sec. \ref{sec:inclination_morphology}).\citetalias{Whitehead2025} reports weak evidence for a mass dependency in $a_2$, though we lack the statistics to comment on this more definitively.
\subsubsection{The regimes of inclination damping}
\label{sec:regimes}
Combining the data from both works, we show $\Delta i/i$ as a function of $i$ in Figure~\ref{fig:di_vs_i_all}. The strength of inclination damping varies significantly across the full inclination range; we identify three rough regions with differentiable behaviour
\begin{itemize}
    \item Exponential Regime $(\sin(i) < 3H_0/R_0)$ \newline
    For objects partially embedded in the disc, inclination decays exponentially with a fixed timescale dependent only on the ambient gas mass. Gas gravity dominates over accretion effects. The gas morphology around the BH is more spherical.
    \item Transitory Regime $(3H_0/R_0 < \sin(i) < 30H_0/R_0)$\newline
    For intermediate inclinations, where the transitor is only very weakly embedded, damping is less efficient for higher inclinations. The gas morphology around the BH is more comet-like. Gas gravity remains the dominant source of damping.
    \item Accretion Regime $(\sin(i)>30H_0/R_0)$\newline
    At very high inclinations, the strength of damping by gravity has fallen sufficiently that accretion effects can begin to contribute more significantly.
\end{itemize}
The geometry of damping by gas differs between the low/high inclination regimes. In the low inclination regime, significant drag can be generated from large scale interactions with the disc, as the transitor perturbs the disc to follow its motion (see Figure 4 of \citetalias{Whitehead2025}). In the high inclination regimes, the high velocity of the transit means that only gas very close to the transitor contributes. Figure~\ref{fig:di_vs_i_all} shows that damping efficiency ($\Delta i / i$) is a monotonic function of inclination $i$; damping is consistently less efficient at higher inclinations, inline with the analytical expectation. This means that for objects initialised at higher inclinations, the majority of their damping lifetime will be spent at these higher inclinations. Given little time is spent at low inclinations, the alignment time can be easily approximated by evolving just Eq. \eqref{eq:p_law_fit}. 
\subsubsection{A combined alignment function}
In Figure \ref{fig:di_vs_i_all} we combine the data from this work with \citetalias{Whitehead2025} and plot the inclination damping as a function of inclination, covering inclinations in the range $i\in[0.1^\circ,15^\circ]$. 
We fit a piecewise powerlaw function to the full data set with the form
\begin{equation}
    \log_{10}\left(|\Delta i|/i\right)_\text{combined} = 
    \begin{cases}
        a_3\log_{10}\left(\tilde{m}\right) + b_3\log_{10}(\tilde{i}_\mathrm{c}) + c_3 & \tilde{i} < \tilde{i}_c\,, \\
        a_3\log_{10}\left(\tilde{m}\right) + b_3\log_{10}(\tilde{i}) + c_3 & \tilde{i} \geq \tilde{i}_c\,,
    \end{cases}
    \label{eq:combined_fit}
\end{equation}
where $\tilde{i}_c$ represents the transition inclination between the exponential low $i$ damping regime (Eq. \ref{eq:henry_fit}) and power law high $i$ regime (Eq. \ref{eq:p_law_fit}). We display the best fit parameters to all three expressions in Table \ref{tab:all_fits}.
\begin{table}
    \centering
    \begin{tabular}{c c c c c c}
    \hline
     Regime                              & $\tilde{i}$ range   &  $X$     & $\mu_X$   & $\sigma_X$ & RMS $\delta$  \\ \hline \hline
     high $i$ (this work)                & $\tilde{i}>3$   & $a_1$             & 0.39      & 0.03 & 0.07\\
                                         &   & $b_1$             & -2.73     & 0.04 & \\
                                         &   & $c_1$             & 1.70      & 0.06 & \\ \hline
     low $i$  (\citetalias{Whitehead2025})& $\tilde{i}<3$ & $a_2$             & 0.69      & 0.03 & 0.20\\
                                         &   & $c_2$             & -0.09     & 0.06 & \\ \hline
     both (combined)               & -  & $a_3$             & 0.67      & 0.02 & 0.18 \\
                                         &   & $b_3$             & -2.64     & 0.06 & \\
                                         &   & $c_3$             & 1.80      & 0.09 & \\ 
                                         &   & $\tilde{i}_c$     & 4.6      & 1.1 & \\ [1ex]
    \hline
    \end{tabular}
    \caption{Best fit parameters $\mu_\text{X}$, their 1-$\sigma$ uncertainties $\sigma_\text{X}$ and fit root-mean-square error for the three models presented in this paper, covering the high inclination regime of this work Eq. \eqref{eq:p_law_fit}, the low inclination regime of \citetalias{Whitehead2025} (Eq. \ref{eq:henry_fit}) and the combined fit of Eq. \eqref{eq:combined_fit} using both datasets. We also show the appropriate inclination ($\tilde{i}=\sin(i)R_0/H_0$) ranges for the individual high and low inclination models.}
    \label{tab:all_fits}
\end{table}

We find the transition inclination between the exponential and transitory damping regime is $\tilde{i}_\text{c}\simeq4.6$. It is not surprising that we would expect a transition regime as the vertical motion of the BH damps towards the scale height of the disc, i.e $\tilde{i}\lesssim 1$, as this marks the approximate inclination where the vertical headwind imparted on the gas around the BH becomes subsonic, i.e $\braket{\dot{z}}/c_\text{s}\lesssim 1$. The true transition $\tilde{i}_\text{c}$ is naturally above unity as the sound speed in the vicinity of the BH is increased. In our low inclination study \citetalias{Whitehead2025}, we explore this transition region of the parameter space in more depth.
\begin{figure*}
    \centering
    \includegraphics[width=17cm]{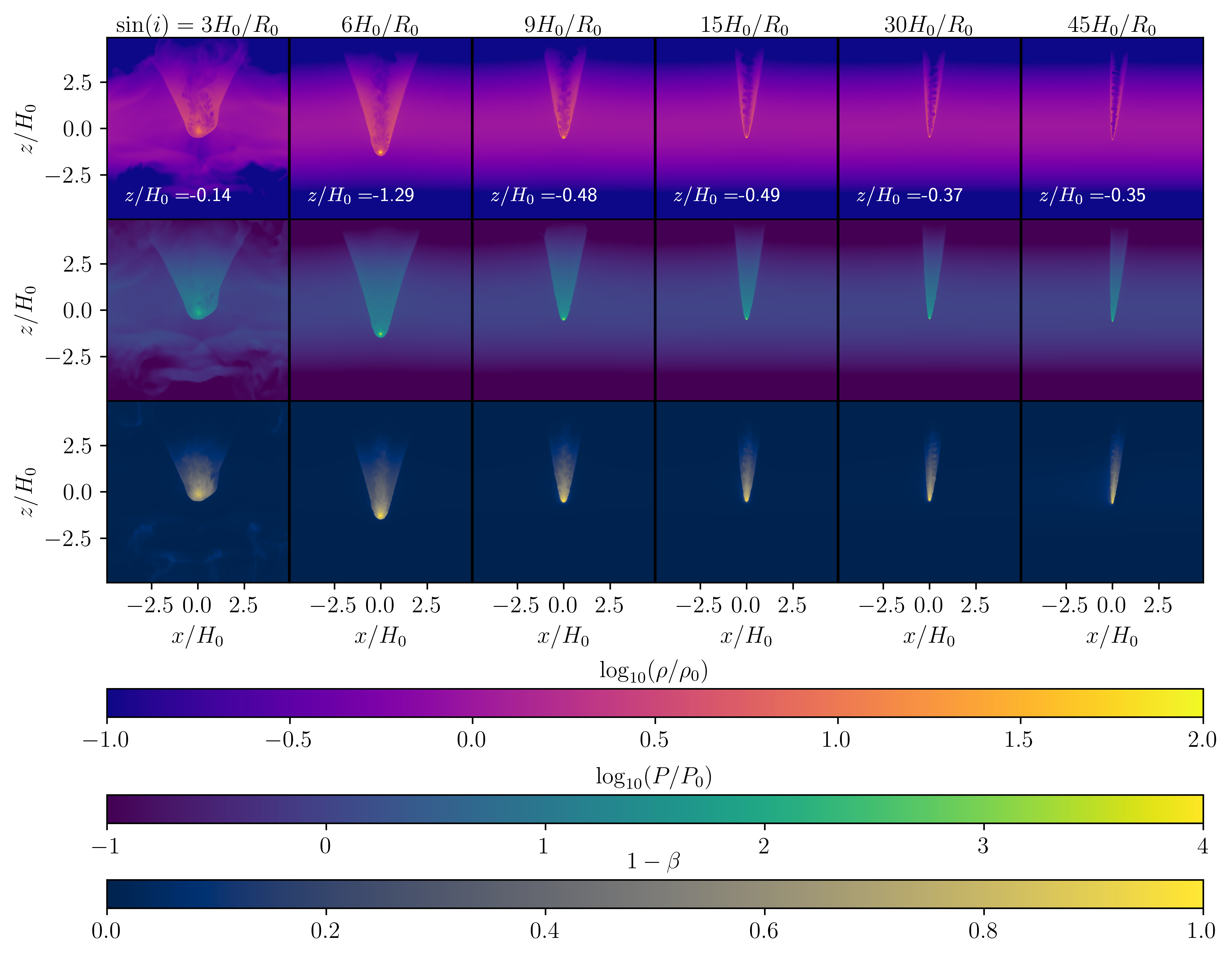}
    \caption{The gas morphology of the wakes generated by transitors across inclinations from $\sin(i)\in[3H_0/R_0,45H_0/R_0]$ and $\Sigma=\Sigma_0$, shown in cross sections at $x=0$. \textit{Top row:} the density $\rho$ relative to the initial midplane density $\rho_0$. \textit{Middle row:} the pressure $P$ relative to the initial midplane pressure $P_0$. \textit{Bottom row:} the ratio of the gas pressure to total pressure $\beta$.  The figure illustrates the increasing width of the wake with declining inclination, until the bow shock undergoes a regime change, becoming more chaotic and spherical for $\sin(i)\lesssim 6H_0/R_0$, coincident with the regime change observed in the inclination damping at $\sin(i)\sim 4.6H_0/R_0$.}
    \label{fig:inclination_morphology}
\end{figure*}

\subsection{Inclination dependence of gas morphology}
\label{sec:inclination_morphology}
To visualise the encounters across our inclination range, we show the gas morphology around the BH as it transitions through the midplane for different inclinations in Figure \ref{fig:inclination_morphology}. The width of the wake is inversely dependent on the inclination, consistent with the higher relative velocity. The underdensity trailing the BH within the wake becomes more pronounced as the gas has less time to respond to the perturbation of the BH as it transits through the disc, driving the reduction in $\varepsilon_\text{grav}$ with $i$. This effect is well documented for simulations of supersonic perturbers in uniform mediums \citep[e.g][]{Canto2013,Grishin2015}. As the inclination of the transitor becomes smaller, the underdense region becomes less pronounced and the mass of the perturbed gas following the BH increases. Meanwhile, the width of the bow shock continues to increase until the bow shock "opens", becoming more spherical and chaotic for inclinations $\sin(i)\lesssim6H_0/R_0$, coincident with the aforementioned change in damping behaviour at $\sin(i)\lesssim(4.6\pm 1)H_0/R_0$. Note this transition is also coincident with the inclination where the accretion cross section $r_\text{c}$ switches from $r_\text{c}\sim R_\text{BHL}$ to $r_\text{c}\sim \rh$ (see Sec. \ref{sec:alignment_times}).  As the BH continues to sink into the disc, the pressure begins to reflect that of a more hydrostatic system, observe the reduced pressure around the BH in the first column of Figure \ref{fig:inclination_morphology} as the ram pressure becomes less significant. Indeed, this is found to be the end state of the low inclination simulations of \citetalias{Whitehead2025}. Radiation pressure dominates in the wake for the majority of our inclination range, only beginning to wane in the same inclination range of $\sin(i)<6H_0/R_0$, although we note $\beta$ retains a value of $\sim0.5$ close to the BH even here.
\section{Discussion}
\label{sec:discussion}
\subsection{Implications}
Population studies of BHs in AGN must account for a vast array of physical processes in order to obtain accurate predictions for present and future GW detections. The predictions for the merging mass function \citep[e.g][]{Yang2019,Rowan2024_rates}, maximum BH mass \citep[e.g][]{Xue2025} and overall rates \citep[e.g][]{Tagawa2020} sensitively depend on the number of BHs embedded in the AGN disc. The embedded population is taken to be either the initial population geometrically within the disc scale height \citep[e.g][]{Delfavero2024}, the number of objects that may embed within the AGN lifetime \citep[e.g][]{Bartos2017,Rowan2024_rates} or modelled in a time-dependent manner based on analytic models (i.e Eqs \ref{eq:t_align_full},\ref{eq:BHL_drag} and \ref{eq:GDF_drag}) for the inclination damping timescale \citep[e.g][]{Tagawa2020,Xue2025}. 

The alignment timescale comparisons of Sec. \ref{sec:alignment_times} imply that the alignment time for moderately inclined objects ($3H_0/R_0 <\sin(i)<30H_0/R_0$) is overestimated by  approximately an order of magnitude. For the studies where the embedded population is simply the objects that embed within $t_\text{AGN}$ (usually taken to be $10^{6}-10^{7}$yr), our findings likely do not have a significant impact on their results as the alignment times form Eq. \ref{eq:t_align_full} and \ref{eq:BHL_drag} also predict that most objects in this inclination range will embed. For studies that allow for a time-dependent disc population \citep[e.g][]{Tagawa2020,Xue2025}, a shorter alignment time would suggest a larger "burst" of BH mergers earlier on in the AGN lifetime \citep[e.g][]{Delfavero2024} as the density of objects in the disc rises faster. Note that with a shorter inclination damping timescale and denser disc population of objects also comes a shorter relaxation timescale, where relaxation from single-single and binary-single encounters competes with the realignment from gas until a quasi-stable velocity dispersion is reached. More efficient disc damping will reduce this characteristic dispersion. Combining this with the low $i$ results from \citetalias{Whitehead2025}, we make an overall prediction that typical encounter energies between isolated BHs and other BHs, or even other binary BH systems are likely lower than previously expected, facilitating easier binary formation \citep[e.g][]{Whitehead2023,Rowan2023,Rowan2024_rates} and with typically low inclinations at formation, i.e  the formed inner binaries have closer to pure prograde $i\sim0^\circ$ or retrograde $i\sim180^\circ$ configurations.
The shock heating generated at shock front leads to a radiation pressure dominated flow around the BH, compared with the gas pressure dominated background. While the BH transits the disc, this radiation is trapped by the high optical depth of the disc. The punctures left by the BH at the exit and entry points of the disc provide an escape point for the breakout emission of this radiation. While we don't perform radiation transfer in this work, semi-analytic studies predict electromagnetic emission with a hardened Wien-like spectrum after each disc transit \citep[e.g][]{Franchini2023,Vurm2025} in what's known as quasi-periodic eruptions or "QPEs". These eruptions are observationally characterised by periodic high amplitude and high energy emissions, detectable via X-ray emission \citep[e.g][]{Acodia2021,chakraborty2024,Arcodia2022,Arcodia2024,Jiang2025,Hernandez2025}, although there is still some uncertainty regarding their astrophysical cause, with others proposing variations in the accretion flow onto the SMBH \citep[e.g][]{Bollimpalli2024,Middleton2025}.

\subsection{Limitations}
This study makes several assumptions and physical simplifications. Below we list considerations not addressed here and their potential implications.
\begin{itemize}
    \item \textit{Viscosity:} The simulations performed here are inviscid. We expect that its effects will primarily manifest in the damping  of the instabilities at the inner edge of the BH wake. As the majority of the drag is caused by gas close to the BH, we don't expect it's inclusion to greatly affect our results.
    \item \textit{Radiative transport:} We have assumed that the gas is always optically thick to radiation so that the gas cannot quickly cool radiatively, allowing us to ignore the need for simulating radiative transport and cooling effects. This assumption will generally hold for the majority of each transit, although one might expect small modifications to the pressure in the underdense centre of the wake behind the BH. A full radiative transport framework would also allow one to make more detailed predictions for observable electromagnetic features of disc transitors.
    \item \textit{Mass accretion:} We have ignored the mass gain of the BH, which may slightly alter the mass flux into our accretion radius. 
    \item \textit{Zero eccentricity:} Our simulations only considered transitors on circular orbits about the AGN. In reality, objects in the NSC have a wide distribution of eccentricities. The relative velocity of the BH to the gas will be eccentricity dependent, hence so will the drag and therefore alignment time. Exploring this parameter in the future will be necessary to better understand the orbital alignment of NSC objects on less idealised (circular) orbits.
    \item \textit{Eddington limited accretion:} We limited the accretion rate onto the BHs to the Eddington Limit. While this is consistent with previous simulations of accretion onto fast moving BHs in a uniform radiation dominated medium, further work is necessary to understand whether this limit is still as robust for our problem. This would require simulating down to the ISCO of the BH, far beyond the resolution limit of our simulations.
    \item \textit{Radiative feedback} - We have not accounted for the radiative feedback from the BH as it accretes. For luminous accretors,  feedback has been shown to reduce the drag on objects moving through a uniform medium \citep[e.g][]{Valesco2019,Valesco2020} (at subsonic velocities at least). Like the accretion, this effect occurs on a subgrid-scale in our treatment, warranting future studies at higher resolution. 
    \item \textit{Restricted parameter space:} We have only considered a small portion of the AGN parameter space, i.e fixed radial position, sound speed and SMBH mass. It is possible that objects may align faster if they are allowed to migrate inwards to denser inner disc regions \citep[e.g][]{Fabj2020}. Further simulations exploring different environmental parameters are necessary to develop a more comprehensive grasp of the alignment time for the full population of inclined objects.
\end{itemize}
\section{Summary and Conclusions}
\label{sec:conclusions}
Motivated to better understand the alignment process NSC objects in AGN, we performed a total of 79 radiative hydrodynamical simulations of BH transits through an varying AGN disc environments with varying inclinations between $i\in[2^\circ,15^\circ]$. We summarise our key findings below:
\begin{itemize}
    \item The transiting BH drives the formation of a strong shock front ahead of the object with a narrow wake trailing its path.
    \item Radiation dominates the thermodynamics of the wake, with $\beta=\frac{P_\text{gas}}{P_\text{gas}+P_\text{rad}}\sim1$ in the vicinity of the BH for the vast majority of the transit.
    \item Gravitation with the wake leads to a net deceleration of the BH in the vertical direction, reducing its inclination. 
    \item The inclination change, relative to the initial inclination ($\Delta i/i$) is a function of inclination itself. We find a power law relation of $\Delta i/i\propto \sin(i)^{-2.7}$, where inclination damping becomes increasingly less efficient at higher inclinations.
    \item For our inclination range, damping is dominated by the BH's gravitation with the wake. As the inclination is increased, gravitational drag becomes less efficient until accretion drag becomes comparable around $\sin(i)\sim30H_0/R_0$. 
    \item We combine the data form this study with our sibling paper examining partially embedded BH satellites \citetalias{Whitehead2025}, developing a comprehensive picture of inclination damping from high to low inclinations, identifying three key regimes (Sec. \ref{sec:low_i_lim}).  
    \item The timescale for the BH to align with the AGN disc is roughly consistent with BHL drag at low inclinations $\sin(i)\sim3H_0/R_0$. However at higher inclinations, we find these analytic predictions overestimate the alignment time by approximately an order of magnitude, which we attribute to the breakdown of the BHL assumption of a uniform and unchanging density and strong radiation pressure.
    \item Our results suggest that more BHs from the NSC may be embedded within the AGN disc than expected from previous analytical expectations, increasing population of potential gas-driven mergers in AGN.
\end{itemize}
In simulating the transit of these compact objects through the AGN disc with more realistic hydrodynamics, we begin to reduce the uncertainty in the alignment time of these objects, a key component in population studies of BH mergers in the AGN channel.

\section*{Acknowledgements}
The research leading to this work was supported by the Independent Research Fund Denmark via grant ID 10.46540/3103-00205B. This work was supported by the Science and Technology Facilities Council Grant Number ST/W000903/1.

\bibliographystyle{mnras}
\bibliography{Paper} 

\appendix

\section{Derivation of alignment time under the impulse approximation.}
\label{app:t_align}
Here we derive the alignment timescale for a BH in an AGN disc assuming the BH receives an impulse from Bondi-Hoyle-Lyttleton accretion as it crosses the disc, as described in \cite{Bartos2017}. The primary assumptions is that the vertical velocity change $\Delta v_\text{z}$ induced by the crossing is directly proportional to the mass accreted $\Delta m_\text{cross}$. Therefore the damping timescale is 
\begin{equation}
    t_\text{align} \sim \frac{T v_\text{z}}{2\Delta v_\text{z}}\sim\frac{T \mBH}{2\Delta m_\text{cross}}
\end{equation}
The crossing mass is given by the cylindrical volume carved out by the BHL radius 
\begin{equation}
  r_\mathrm{BHL}=2G\mBH/(\Delta v^{2}+c_s^{2})\,,  
\end{equation}
where 
\begin{equation}
    \Delta v=v_\mathrm{orb}((1-\cos(i))^{2}+\sin^{2}(i))^{1/2}=2v_\mathrm{orb}\sin\bigg(\frac{i}{2}\bigg)
\end{equation}
is the relative velocity of the BH to the gas, which orbits the SMBH with velocity $v_\mathrm{orb}=\sqrt{G\mSMBH/R}$. The crossing mass is then
\begin{equation}
    \Delta m_\mathrm{cross}=\Delta v t_\mathrm{cross}r_\mathrm{BHL}^{2}\pi\Sigma/(2H_0) 
\end{equation}
with the crossing time given by
\begin{equation}
    t_\mathrm{cross}\approx2H_0/(v_\mathrm{orb}\sin{i}) \,, 
\end{equation}
where we have assumed the disc has a thickness $2H_0$ and uniform density $\Sigma/(2H_0)$.
Putting all this together gives an alignment time of 
\begin{equation}
t_\mathrm{align}=\frac{t_\mathrm{orb}}{2}\frac{\cos(i/2)(\Delta v^{2}+c_s^{2})^{2}}{4G^{2}\mBH\pi\Sigma}
\end{equation}
where the identity $\sin(i)/\sin(i/2)=2\cos(i/2)$ has been applied.
% do not change these lines
%\bsp	% typesetting comment
\label{lastpage}
\end{document}